\documentclass[twocolumn]{revtex4}
\usepackage{epsfig}
\usepackage{amssymb}
\usepackage{amsmath}
\usepackage{amsfonts}
\usepackage{graphicx}
\usepackage{mathrsfs}
\usepackage[dvips]{color}
\usepackage{multirow}

\newcommand{\R}{\mathbb{R}}
\newcommand{\C}{\mathbb{C}}

\newcommand{\fg}{\mathfrak{g}}

\newcommand{\fz}{\mathfrak{z}}

\newcommand{\fK}{\mathfrak{K}}

\newcommand{\bM}{\mathbf{M}}

\newcommand{\cF}{\mathcal{F}}

\newcommand{\cP}{\mathcal{P}}

\newcommand{\cT}{\mathcal{T}}

\newcommand{\be}{\begin{equation}}
\newcommand{\ee}{\end{equation}}
\newcommand{\bea}{\begin{eqnarray}}
\newcommand{\eea}{\end{eqnarray}}
\newcommand{\nn}{\nonumber}

\newcommand{\ed}{\end{document}}

\newcommand{\rl}{{\rm l}}
\newcommand{\rr}{{\rm r}}

\newcommand{\bi}{\begin{itemize}}
\newcommand{\ei}{\end{itemize}}

\newcommand{\bce}{\begin{center}}
\newcommand{\ece}{\end{center}}

\newcommand{\RE}{{\rm Re}}
\newcommand{\IM}{{\rm Im}}

\begin{document}

\title{Nonlinear Scattering and Its Transfer Matrix Formulation in
One Dimension}

\author{Ali~Mostafazadeh\\[6pt]
Departments of Mathematics and Physics, Ko\c{c} University,\\
34450 Sar{\i}yer, Istanbul, Turkey}

\begin{abstract}

We present a systematic formulation of scattering theory for
nonlinear interactions in one dimension and develop a
nonlinear generalization of the transfer matrix that has
a composition property similar to its linear analog's. We offer
alternative characterizations of spectral singularities, unidirectional
reflectionlessness and invisibility, and nonreciprocal transmission
for nonlinear scattering systems, and examine the application of
our general results in addressing the scattering problem for
nonlinear single- and double-$\delta$-function potentials.

\medskip

\end{abstract}

\maketitle

\section{Introduction}

If the strength of a nonlinear interaction has a sufficiently rapid asymptotic decay rate its scattering effects can be studied using the framework offered by standard quantum scattering theory. For example, consider a time-harmonic scalar wave, $e^{-i\omega t}\psi(x)$, whose interaction with a scatterer is described by the wave equation
    \be
    -\psi''(x)+\cF(x,\psi)=\fK^2\psi(x),
    \label{NLSE}
    \ee
where $\psi(x)$ is a possibly complex-valued wave function, a prime stands for differentiation with respect to $x$, and $\cF(x,\psi)$ represents the interaction. The well-known time-independent potential scattering \cite{newton} corresponds to situations where $\cF(x,\psi)$ is a linear function of $\psi$, i.e., there is a potential $v(x)$ such that $\cF(x,\psi)=v(x)\psi(x)$. We can also consider scattering by a nonlocal potential \cite{muga} which
corresponds to $\cF(x,\psi)=\int_{-\infty}^\infty v(x,x')\psi(x')dx'$.

Suppose that as $x\to\pm\infty$ the function $\cF(x,\psi)$ decays to zero at such a rate that the global solutions of (\ref{NLSE}) satisfy the following asymptotic boundary conditions
    \bea
    \psi(x)&\to& A_- e^{i\fK x}+B_- e^{-i\fK x}~~~{\rm for}~~~x\to-\infty,
    \label{ini-m}\\
    \psi(x)&\to& A_+ e^{i\fK x}+B_+ e^{-i\fK x}~~~{\rm for}~~~x\to\infty,
    \label{ini-p}
    \eea
where $A_\pm$ and $B_\pm$ are complex coefficients. In this case, we can introduce the scattering (Jost) solutions, $\psi_{l}$ and $\psi_{r}$, of (\ref{NLSE}) that respectively correspond to setting $B_+=0$ and $A_-=0$. $\psi_{l/r}$ describes the scattering of an incident wave that is sent from the left/right of the region in which the interaction has sizeable strength. If we denote the complex amplitude of the incident wave by $A^{l/r}$, we can identify $\psi_{l/r}$ with solutions satisfying:
    \bea
    \psi_{l}(x)&\to&\left\{\begin{array}{ccc}
    A^{l} \left(e^{i\fK x}+R^{l}e^{-i\fK x}\right) & {\rm for} & x\to-\infty,\\
    A^{l} T^{l}e^{i\fK x} & {\rm for} & x\to+\infty,\end{array}\right.~~~
    \label{psi-left}\\
    \psi_{r}(x)&\to&\left\{\begin{array}{ccc}
    A^{r}T^{r}e^{-i\fK x}& {\rm for} & x\to-\infty,\\
    A^{r} \left(e^{-i\fK x}+R^{r}e^{i\fK x}\right)  &
    {\rm for} & x\to+\infty,\end{array}\right.
    \label{psi-right}
    \eea
where $R^{l/r}$ and $T^{l/r}$ are respectively the left/right reflection and transmission amplitudes \cite{prl-2013}. If the interaction is nonlinear these are complex-valued functions of $\fK$, but in general they depend on both $\fK$ and $A^{l/r}$. The ultimate aim of solving the scattering problem for a given interaction is to determine the nature of the dependence of $R^{l/r}$ and $T^{l/r}$ on $\fK$ and $A^{l/r}$.

Conditions (\ref{ini-m}) and  (\ref{ini-p}) are  certainly satisfied
for finite-range interactions. These are interactions that are
confined to a finite interval $[a,b]$, i.e., $\cF(x,\psi)=0$ for
$x\notin[a,b]$. For such interactions we can replace $x\to-\infty$
and $x\to+\infty$ in (\ref{ini-m}) -- (\ref{psi-right}) by $x\leq a$
and $x\geq b$, respectively. In the following we confine our
attention to finite-range interactions that vanish outside $[a,b]$.
The study of more general interactions corresponds to taking
$a\to-\infty$ and $b\to\infty$ limit of the results for finite-range
interactions. This may require dealing with difficult mathematical
problems \cite{NL-scattering} whose discussion is beyond the scope
of the present investigation.

This article is organized as follows. In Sec.~\ref{S2} we  offer a
systematic formulation of the scattering problem for finite-range
nonlinear interactions. In Sec.~\ref{S3}, we apply this formulation
to solve the scattering problem for a class of nonlinear
$\delta$-function potentials. In Sec.~\ref{S4}, we devise an
alternative transfer-matrix formulation of the scattering theory and
explore its basic properties. In Sec.~\ref{S5} we determine the
nonlinear transfer matrix for a class of nonlinear double
$\delta$-function potentials, and in Sec.~\ref{S6} we summarize our
main findings and present our concluding remarks.

\section{Solution of the scattering problem using Jost functions}
\label{S2}

In view of (\ref{psi-left}) and (\ref{psi-right}), we can respectively identify $\psi_{l}$ and $\psi_{r}$ with the solution of the initial-value problem defined by (\ref{NLSE}) and the initial conditions \cite{prl-2013}:
    \begin{align}
    &\psi_{l}(b)=N_{l} e^{ib\fK}, && \psi_{l}'(b)=i\fK N_{l}e^{ib\fK},
    \label{ini-1}\\
    &\psi_{r}(a)=N_{r}e^{-ia\fK}, && \psi_{r}'(a)=-i\fK N_{r}e^{-ia\fK},
    \label{ini-2}
    \end{align}
in the interval $[a,b]$, where
    \be
    N_{l/r}:=A^{l/r}T^{l/r}.
    \ee

The determination of $R^{l/r}$ and $T^{l/r}$ turns out to involve the following quantities that are called Jost functions:
    \be
    \begin{aligned}
    &F^{l/r}_\pm:=\psi'_{l/r}(b)\pm i\fK \psi_{l/r}(b),\\
    & G^{l/r}_\pm:=\psi'_{l/r}(a)\pm i\fK \psi_{l/r}(a).
    \end{aligned}
    \label{F-G}
    \ee
Replacing $x\to-\infty$ and $x\to+\infty$ in (\ref{ini-m}) -- (\ref{psi-right}) by $x\leq a$ and $x\geq b$, and using the resulting relations together with (\ref{ini-1}) -- (\ref{F-G}), we find
    \begin{align}
    &G_+^{l}=2i\fK\, e^{ia\fK}A^{l},
    && F_-^{r}=-2i\fK\, e^{-ib\fK}A^{r},
    \label{eq1}\\
    & R^{l}=-\frac{ e^{2ia\fK} G_-^{l}}{G^{l}_+},
    &&R^{r}=-\frac{e^{-2ib\fK}F_+^{r}}{F^{r}_-},
    \label{Rs}\\
    & T^{l}=\frac{e^{i(a-b)\fK}F^{l}_+}{G^{l}_+},
    &&T^{r}=\frac{e^{i(a-b)\fK}G_-^{r}}{F^{r}_-}.
    \label{Ts}
    \end{align}
These equations provide a basic framework for the solution of the scattering problem we consider in this article. More specifically, they suggest that we proceed as follows:
    \begin{enumerate}
    \item Find the solutions $\psi_{l/r}$ of (\ref{NLSE}) in $[a,b]$
    that fulfil the initial conditions (\ref{ini-1}) and (\ref{ini-2}).
    \item Determine the Jost functions: $F_+^{l/r}$, $F_-^{r}$, $G_+^{l}$, and $G_-^{l/r}$.
    \item Substitute the expression for $F_+^{l/r}$ and $G_-^{l/r}$
    in (\ref{Rs}) and (\ref{Ts}) to obtain $R^{l/r}$ and
    $T^{l/r}$ as functions of $N_{l/r}$.
    \item Substitute the expression for $F_-^{r}$ and $G_+^{l}$ in
    (\ref{eq1}) and solve the resulting equations for $N_{l/r}$. This
    determines $N_{l/r}$ as a (possibly multivalued) function of $A^{l/r}$ and $\fK$.
    \item Use the results of steps 3 and 4 to express $R^{l/r}$ and
    $T^{l/r}$ in terms of $A^{l/r}$ and $\fK$.
    \end{enumerate}
This prescription applies to the scattering problems for any linear or nonlinear interaction that is confined to a closed interval $[a,b]$ in space. If the interaction is linear, there is no need for steps 4 and 5, because step~3 yields expressions for $R^{l/r}$ and $T^{l/r}$ that are independent of $N^{l/r}$. The main difficulty in dealing with nonlinear scattering problems is therefore associated with step 4.

We can use Eqs.~(\ref{eq1}) -- (\ref{Ts}) to obtain a characterization of the phenomena of spectral singularities \cite{prl-2009,prl-2013}, unidirectional reflectionlessness, transparency, and invisibility \cite{lin,unidir1,pra-2013,pla-2017}, and nonreciprocal transmission \cite{diode} in terms of the Jost functions. Table~\ref{table1} summarizes the outcome.
    \begin{table}[!htbp]
    \begin{center}
    \begin{tabular}{|c|c|c|}
    \hline
    Phenomenon & Definition &  Characterization\\
    \hline
    Spectral Singularity ($l$) & $R^{l}\!=\! T^{l}\!=\! \infty$ & $G^{l}_+=0$\\
    \hline
    Spectral Singularity ($r$) & $R^{r}\!=\!T^{r}\!=\!\infty$ & $F^{r}_-=0$\\
    \hline
    Reflectionlessness ($l$) & $R^{l}=0$ & $G^{l}_-=0$\\
    \hline
    Reflectionlessness ($r$) & $R^{r}=0$ & $F^{r}_+=0$\\
    \hline
    Transparency ($l$) & $T^{l}=1$ & $F^{l}_+=e^{i(b-a)\fK}G^{l}_+$\\
    \hline
    Transparency ($r$) & $T^{r}=1$ & $G^{r}_-=e^{i(b-a)\fK}F^{r}_-$\\
    \hline
    \begin{tabular}{c}
    Nonreciprocal\\
    Transmission
    \end{tabular}
    &
    \begin{tabular}{c}
    $T^{l}\neq T^{r}$\\
    for $A^{l}=A^{r}$
    \end{tabular}
    &
    \begin{tabular}{c}
    $F^{l}_+\neq -e^{i(a+b)\fK}G^{r}_-$\\
    for $A^{l}=A^{r}$
    \end{tabular}\\
    \hline
    \end{tabular}
    \caption{Characterization of various scattering phenomena in terms of the Jost functions $F^{{l/r}}_\pm$ and $G^{{l/r}}_\pm$. ``($l$)'' and ``($r$)'' respectively abbreviate ``from left'' and ``from right''. The equations given in the third column restrict
    the allowed wavenumber $\fK$, which takes a real and positive value, and the complex
    amplitude of the incident wave, $A^{{l/r}}$. In principle they can be solved
    to express $A^{{l/r}}$ as a function of $\fK$.}
    \label{table1}
    \end{center}
    \end{table}

\section{Nonlinear $\delta$-function potentials}
\label{S3}

As a simple example consider the nonlinear point interaction defined by
    \be
    \begin{aligned}
    &\cF(x,\psi):=v(x,\psi)\psi,\\
    &v(x,\psi):=f(|\psi(x)|)\delta(x-c),
    \end{aligned}
    \label{pt}
    \ee
where $f:\R\to\C$ is continuous function. Clearly this is a nonlinear point interaction if and only if $f$ is a non-constant function. A well-known choice for $f$ is
    \be
    f(|\psi|)=\fz|\psi|^\nu,
    \label{f-power}
    \ee
where $\fz$ is a possibly complex nonzero coupling constant, and $\nu$ is a real number greater than $-1$. The special cases where $\fz$ takes a negative real value is studied in \cite{molina}. For $\nu=2$, (\ref{f-power}) defines a localized Kerr nonlinearly \cite{malomed}. For $\nu=0$, it corresponds to a $\delta$-function potential with a possibly complex coupling constant \cite{jpa-2016b}.

Because the point interaction (\ref{pt}) vanishes for $x\neq c$, we can take
    \begin{align}
    &a:=c-\epsilon,
    &&b:=c+\epsilon,
    \label{ab-pt}
    \end{align}
for any $\epsilon>0$. The solution of the nonlinear Schr\"odinger equation (\ref{NLSE}) for (\ref{pt}) is straightforward. It is given by
    \be
    \psi(x)=\left\{\begin{array}{ccc}
    A_- e^{i\fK x}+B_- e^{-i\fK x}~~~{\rm for}~~~x<c,\\
    A_+ e^{i\fK x}+B_+ e^{-i\fK x}~~~{\rm for}~~~x>c,
    \end{array}\right.
    \label{sol=pt}
    \ee
and the matching conditions
    \be
    \begin{aligned}
    &\psi(c^+)=\psi(c^-)=\psi(c),\\
    &\psi'(c^+)=\psi'(c^-)+f(|\psi(c)|)\,\psi(c),
    \end{aligned}
    \label{match}
    \ee
where for any function $\phi:\R\to\C$, $\phi(c^+)$ and $\phi(c^-)$
respectively stands for the right and left limit of $\phi(x)$ at
$x=c$.

The Jost solutions $\psi_{l}$ and $\psi_{r}$ are solutions of (\ref{NLSE}) that fulfil the initial conditions (\ref{ini-1}) and (\ref{ini-2}), respectively. Comparing (\ref{psi-left}) and (\ref{sol=pt}), we see that for $\psi=\psi_{l}$,
    \begin{align}
    &A_+=N_{l},
    &&B_+=0.
    \label{left-x1}
    \end{align}
Substituting these in (\ref{sol=pt}) and enforcing (\ref{match}), we find
    \begin{align}
    &A_-=N_{l}\left(1+\frac{if^{l}}{2\fK}\right),
    &&B_-=-\frac{if^l N_{l} e^{2ic\fK}}{2\fK},
    \label{left-x2}
    \end{align}
where
    \[f^l:=f(|N_l|).\]
Eqs.~(\ref{sol=pt}), (\ref{left-x1}), and (\ref{left-x2}) show that
    \be
    \psi_{l}(x)=N_{l} e^{i\fK x}\times
    \left\{\begin{array}{ccc}
    \phi_{l}(x) &{\rm for}&x<c,\\
    1&{\rm for}&x\geq c,
    \end{array}\right.
    \label{psi-left-pt}
    \ee
where
    \be
    \phi_{l}(x):=1+\frac{i f^{l}[1-e^{2i\fK (c-x)}]}{2\fK}.
    \nn
    \ee
A similar analysis yields
    \be
    \psi_{r}(x)=N_{r} e^{-i\fK x}\times
    \left\{\begin{array}{ccc}
    1&{\rm for}&x\leq c,\\
    \phi_{r}(x)&{\rm for}&x>c,
    \end{array}\right.
    \label{psi-right-pt}
    \ee
where
    \begin{align*}
    &\phi_{r}(x):=1+\frac{if^r[1-e^{2i\fK (x-c)}]}{2\fK},
    &&f^r:=f(|N_r|).
    \nn
    \end{align*}

Next, we use (\ref{psi-left-pt}) and (\ref{psi-right-pt}) to determine the Jost functions~(\ref{F-G})
with $a$ and $b$ given by (\ref{ab-pt}) in the limit $\epsilon\to 0$. This results in
    \begin{align}
    &F_+^{l}=2i\fK N_{l} e^{ic\fK},\quad\quad\quad G_-^{l}=-f^{l} N_{l} e^{ic\fK},\\
    \label{FpL}
    &G_+^{l}=2i\fK N_{l} e^{ic\fK}\left(1+\frac{if^{l}}{2\fK}\right),\\
    &F_-^{r}=-2i\fK N_{r} e^{-ic\fK}\left(1+\frac{if^r}{2\fK}\right),\\
    &F_+^{r}=f^r N_{r} e^{-ic\fK},\quad\quad\quad
    G_-^{r}=-2i\fK N_{r} e^{-ic\fK}.
    \label{GmR}
    \end{align}
Substituting these in (\ref{eq1}) -- (\ref{Ts}) and simplifying the result, we find
    \begin{align}
    &A^{l/r}=N_{l/r}\left(1+\frac{if^{l/r}}{2\fK}\right),
    \label{As=}\\
    &R^{l}=\frac{-i e^{2ic\fK}f^{l}}{2\fK  +if^{l}},
    &&T^{l}=\frac{2\fK}{2\fK+if^{l}},
    \label{left-RT}\\
    &R^{r}=\frac{-i e^{-2ic\fK}f^r}{2\fK  +if^r},
    &&T^{r}=\frac{2\fK}{2\fK+if^{r}}.
    \label{right-RT}
    \end{align}
According to these equations, the transformation $\rl\leftrightarrow\rr$ implies $R^{l}\leftrightarrow e^{-4ic\fK}R^{r}$ and  $T^{l}\leftrightarrow T^{r}$. This transformation rule has its root in the invariance of the nonlinear $\delta$-function potential (\ref{pt}) under the reflection: $x\to 2c-x$. For a general discussion of the effect of reflection on scattering data, see \cite{bookchapter}.

The final step in solving the scattering problem for the interaction (\ref{pt}) is to solve (\ref{As=}) for $N_{{l/r}}$ in terms of $A^{{l/r}}$ and substitute the result in (\ref{left-RT}) and (\ref{right-RT}). To this end, we evaluate the absolute value of both side of (\ref{As=}) to show that $|N_{l/r}|$ is a solution of
    \begin{align}
    &x^2|\hat f(x)|^2\,-2x^2\IM[\hat f(x)]+x^2-|A^{{l/r}}|^2=0,
    \label{eqn-pt}
    \end{align}
where $\hat f(x):=f(x)/2\fK$. We can determine the reflection and transmission amplitudes by finding a real and nonnegative solution of this equation, identifying it with $|N_{l/r}|$, and setting $f^{l/r}=f(|N_{l/r}|)$ in (\ref{left-RT}) and  (\ref{right-RT}).

For example, for a localized Kerr nonlinearity, where $f(x)=\fz\, x^2$, Eq.~(\ref{eqn-pt}) takes the form:
    \be
    |\hat\fz|^2x^6-2\IM(\hat\fz)x^4+x^2-|A^{l/r}|^2=0,
    \ee
where $\hat\fz=\fz/2k$. This is a cubic equation in $x^2$ with explicit formulas for its solutions. It is not difficult to show that only one of these solutions is real and positive. This show that $|N_{l/r}|$ is a single-valued function of $|A^{l/r}|$, the reflection and transmission amplitudes are uniquely determined, and bistabilities and multistabilities are absent. This is not the case if $f(x)$ is a quadratic polynomial with a nonzero linear term \cite{malomed}.

According to (\ref{As=}) and (\ref{left-RT}), whenever $f^{l}$ takes imaginary values with a positive imaginary part so that
    \be
    \fK_\star:=-\frac{if^{l}}{2}=-\frac{f(|N_l|)}{2}
    \label{K-ss}
    \ee
is real and positive, $R^{l}$ and $T^{l}$ blow up at $\fK=\fK_\star$ while $A^{l}=0$. This marks a (nonlinear) spectral singularity \cite{pra-2013,jo-2017} at which the nonlinear $\delta$-function potential (\ref{pt}) emits outgoing radiation to the right and the left of $x=c$. The intensity of the wave reaching $x=-\infty$ and $x=+\infty$, which are respectively given by $I^{l}_-:=|A^{l} R^{l}|^2$ and $I^{l}_+:=|A^{l} T^{l}|^2$  turn out to coincide; $I^{l}_\pm=|N_{l}|^2$, where $|N_l|$ is such that  (\ref{K-ss}) holds.

If the interaction models an optical system, the emergence of a
spectral singularity corresponds to fulfilling the laser threshold
condition \cite{pra-2011,jo-2017}. The conditions $\RE(f^{l})=0$ and
$\IM(f^{l})>0$ shows that (\ref{pt}) represents a thin film made of a
nonlinear high-gain material \cite{gadallah}. According
to our analysis such a film should begin lasing with output
intensity $I^{l}_\pm$ at the wavenumber (\ref{K-ss}). If we reverse
the sign of $f^{l}$, the system supports a time-reversed spectral
singularity \cite{jpa-2012} and functions as a nonlinear coherent perfect
absorber \cite{CPA1a,CPA1b,CPA2,CPA3}.

Next, suppose that $f(x)$ has a real and positive zero $x_\star$,
and consider the scattering of a left/right incident such that
$|A^{l/r}|=x_\star$. Then we can satisfy (\ref{eqn-pt}) by setting
$x=x_\star$. Furthermore, because
$f^{l/r}=f(|N_{l/r}|)=f(x_\star)=0$, Eqs.~(\ref{left-RT}) and
(\ref{right-RT}) imply $R^{l/r}=T^{l/r}-1=0$. Therefore, the point
interaction~(\ref{pt}) is reflectionless and transparent (and hence
invisible) both from the left and right. This can also be seen
directly from Table~\ref{table1}, because according to (\ref{FpL})
and (\ref{GmR}), $f^{l/r}=0$ implies $F^r_+=G^l_-=0$, $F^l_+=G^l_+$,
and $F^r_-=G^r_-$.

\section{Nonlinear Transfer matrix}
\label{S4}

For a linear scattering problem the transfer matrix $\bM$ is defined
to be the $2\times 2$ matrix connecting the coefficients $A_\pm$ and
$B_\pm$ of (\ref{ini-m}) and (\ref{ini-p})according to
\cite{razavy,prl-2009,sanchez}:
    \be
    \left[\begin{array}{cc}
    A_+\\B_+\end{array}\right]=\bM\left[\begin{array}{cc}
    A_-\\B_-\end{array}\right].
    \label{M=}
    \ee
This equation determines $\bM$ uniquely, if we demand that it does not depend on $A_-$ and $B_-$. We propose to use (\ref{M=}) as the definition of the transfer matrix also for the nonlinear scattering problems. The existence of such a matrix follows from that of global scattering solutions satisfying the asymptotic boundary conditions (\ref{ini-m}) and (\ref{ini-p}). Its uniqueness however cannot be ensured, because the nonlinear nature of the corresponding wave equation makes every matrix $\bM$ fulfilling (\ref{M=}) depend on $A_-$ or $B_-$.

It is not difficult to characterize the non-uniqueness of the transfer matrix in terms of a pair of arbitrary continuous functions $f_1,f_2:\C^2\to\C$; it is easy to check that if $\bM$ satisfies (\ref{M=}), then so does $\bM+\delta\bM$, where
    \[ \delta\bM:=\left[\begin{array}{cc}
    f_1(A_-,B_-)B_- & -f_1(A_-,B_-)A_-\\
    f_2(A_-,B_-)B_- & -f_2(A_-,B_-)A_-\end{array}\right].\]
Surprisingly, however, the lack of uniqueness of $\bM$ does not obstruct its application in solving nonlinear scattering problems, for in view of (\ref{psi-left}), (\ref{psi-right}), and (\ref{M=}), we have
    \begin{align}
    &R^{l}=-\frac{M_{21}^{l}}{M_{22}^{l}},
    &&T^{l}=\frac{\det\bM^{l}}{M_{22}^{l}},
    \label{RT-left}\\
    &R^{r}=\frac{M_{12}^{r}}{M_{22}^{r}},
    &&T^{r}=\frac{1}{M_{22}^{r}},
    \label{RT-right}
    \end{align}
where $M_{ij}^{l/r}$ are the entries of
    \begin{align}
    &\bM^{l}:=\bM(A^{l},A^{l} R^{l}),
    &&\bM^{r}:=\bM(0,A^{r} T^{r}).
    \label{M-left-right}
    \end{align}
Under the transformation $\bM\to\bM+\delta\bM$, $\bM^r$ is left invariant. Therefore this transformation does not affect the right-hand side of  (\ref{RT-right}). Substituting $M_{ij}^l+\delta M_{ij}^l$ for $M^l_{ij}$ in (\ref{RT-left}), we have shown that these equation are also invariant.

Eqs.~(\ref{RT-left}) and (\ref{RT-right}) provide a simple characterization of spectral singularities, reflectionlessness, transparency, and nonreciprocal transmission in terms of entries of the nonlinear transfer matrix. Table~\ref{table2} summarizes this characterization scheme.
\begin{table}[!htbp]
    \begin{center}
    \begin{tabular}{|c|c|c|}
    \hline
    Phenomenon & Definition &  Characterization\\
    \hline
    Spectral Singularity ($l$) & $R^{l}\!=\! T^{l}\!=\! \infty$ & $M_{22}^l=0$\\
    \hline
    Spectral Singularity ($r$) & $R^{r}\!=\!T^{r}\!=\!\infty$ & $M_{22}^r=0$\\
    \hline
    Reflectionlessness ($l$) & $R^{l}=0$ & $M_{21}^l=0$\\
    \hline
    Reflectionlessness ($r$) & $R^{r}=0$ & $M_{12}^r=0$\\
    \hline
    Transparency ($l$) & $T^{l}=1$ & $M_{22}^l=\det\bM^l$\\
    \hline
    Transparency ($r$) & $T^{r}=1$ & $M_{22}^r=1$\\
    \hline
    \begin{tabular}{c}
    Nonreciprocal\\
    Transmission
    \end{tabular}
    &
    \begin{tabular}{c}
    $T^{l}\neq T^{r}$\\
    for $A^{l}=A^{r}$
    \end{tabular}
    &
    \begin{tabular}{c}
    $M_{22}^l\neq M_{22}^r\det\bM^l$\\
    for $A^{l}=A^{r}$
    \end{tabular}\\
    \hline
    \end{tabular}
    \caption{Characterization of various scattering phenomena in terms of the entries of nonlinear transfer matrix.}
    \label{table2}
    \end{center}
    \end{table}

In practice, we can find $\bM(A_-,B_-)$ for  arbitrary choices of
$A_-$ and $B_-$ by solving the initial-value problem defined by
(\ref{NLSE}) and (\ref{ini-m}) and imposing (\ref{M=}). This
determines $\bM(A_-,B_-)$ up to the choice of the functions $f_1$
and $f_2$. We can choose these functions arbitrarily, because they
do not enter in (\ref{RT-left}) and (\ref{RT-right}). In view of
(\ref{M-left-right}), these relations provide two pairs of complex
equations that we can, in principle, solve to express
$(R^{l},T^{l})$ and $(R^{r},T^{r})$ in terms of $(\fK,A^{l})$ and
$(\fK,A^{r})$, respectively. In general these equations may have
more than one solution. This marks the emergence of bistable and
multistable reflection and transmission profiles
\cite{moya,wang-2008}.

Transfer matrices provide an indispensable tool for dealing with locally periodic linear scattering problems \cite{yeh,pereyra,griffiths} because of their composition property \cite{sanchez,ap-2015,bookchapter}. Nonlinear transfer matrices we have introduced in this article possess a similar composition property. To see this, suppose that there is a real number $c$ such that we can decompose the interaction term $\cF(x,\psi)$ in (\ref{NLSE}) into the sum of two separate parts, i.e.,
    \be
    \cF(x,\psi)=\cF_1(x,\psi)+\cF_2(x,\psi),
    \label{F=FF}
    \ee
where $\cF_1(x,\psi)=0$ for $x>c$ and $\cF_2(x,\psi)=0$ for $x<c$. Then we can use (\ref{M=}) to show that the transfer matrix $\bM^{(j)}$ associated with the interaction $\cF_j(x,\psi)$, with $j=1,2$, satisfies the composition rule:
    \be
    \bM^{(2)}(A_0,B_0)\bM^{(1)}(A_-,B_-)=\bM(A_-,B_-),
    \label{compose}
    \ee
where
    \be
    \left[\begin{array}{c}A_0\\B_0\end{array}\right] :=\bM^{(1)}(A_-,B_-)
    \left[\begin{array}{c}A_-\\B_-\end{array}\right].
    \label{AB-zero}
    \ee
We abbreviate (\ref{compose}) as
    \be
    \bM^{(2)}\circ\bM^{(1)}=\bM.
    \label{compose-short}
    \ee
If $\cF_1(x,\psi)$ and $\cF_2(x,\psi)$ have a linear dependence on $\psi$, $\bM^{(1)}$ and $\bM^{(2)}$ depend only on $\fK$,  $\bM^{(2)}\circ\bM^{(1)}=\bM^{(2)}\bM^{(1)}$, and (\ref{compose-short}) reduces to the usual composition rule for linear transfer matrices  \cite{sanchez,ap-2015}, namely $\bM=\bM^{(2)}\bM^{(1)}$. We can prove (\ref{compose}) using the same argument one usually employs in the proof of its linear analog \cite{bookchapter}.

The determination of the nonlinear transfer matrix for the nonlinear $\delta$-function potentials (\ref{pt}) is straightforward. We can use the matching conditions (\ref{match}) to express $A_+$ and $B_+$ appearing in (\ref{sol=pt}) in terms of $A_-$ and $B_-$. This gives
    \bea
    A_+&=&\left(1- {\fg} \right)A_- - {\fg}\, e^{-2ic\fK}B_- ,
    \label{Ap=}\\
    B_+&=& {\fg}\, e^{2ic\fK}A_- +\left(1+ {\fg} \right)B_- ,
    \label{Bp=}
    \eea
where
    \be
    {\fg}:=\frac{i}{2k}\left[f(|e^{2ic\fK}A_-+B_-|)\right].
    \label{f-minus}
    \ee
Eqs.~(\ref{M=}), (\ref{Ap=}), and (\ref{Bp=}) suggest that we take
    \be
    \bM(A_-,B_-)=\left[\begin{array}{cc}
    1- {\fg} & - {\fg}\,e^{-2ic\fK}\\
    {\fg}\,e^{2ic\fK}&1+ {\fg}\end{array}\right].
    \label{M-pt}
    \ee
This in particular implies $\det\bM=1$. Hence $\det\bM^{l}=1$.

In view of (\ref{RT-left}) and (\ref{RT-right}) and the fact that $\det\bM^{l}=1$, we can obtain the transmission amplitudes using the nonlinear transfer matrix (\ref{M-pt}) provided that we calculate $M_{21}^{l}$, $M_{22}^{l}$, $M_{12}^{r}$, and $M_{22}^{r}$. We do this using (\ref{M-left-right}) and (\ref{M-pt}). Substituting the result in (\ref{RT-left}) and (\ref{RT-right}), we then find
    \begin{align}
    &R^{l}=\frac{-{\fg}^l\, e^{2ic\fK}}{1+{\fg}^l},
    &&T^{l}=\frac{1}{1+{\fg}^l},
    \label{RT-left-pt}\\
    &R^{r}=\frac{-{\fg}^r\, e^{-2ic\fK}}{1+{\fg}^r},
    &&T^{r}=\frac{1}{1+{\fg}^r},
    \label{RT-right-pt}
    \end{align}
where
    \bea
    {\fg}^{l}&:=&\frac{i}{2k}f\big(|A^le^{i\fK c}+e^{-i\fK c}A^lR^l|\big),\nn\\
    {\fg}^{r}&:=&\frac{i}{2k}f\big(|A^rT^r|\big).\nn
    \eea
According to (\ref{RT-left-pt}), $|A^le^{i\fK c}+e^{-i\fK c}A^lR^l|=
|A^l||e^{-2i\fK c}R^l+1|=|A^lT^l|$. We also recall that $A^{l/r}T^{l/r}=N_{l/r}$. These observations show that
    \begin{align}
    &{\fg}^l=\frac{i f\big(|N_l|\big)}{2\fK}=\frac{if^{l}}{2\fK},
    &&{\fg}^r=\frac{i f\big(|N_r|\big)}{2\fK}=\frac{if^r}{2\fK}.
    \end{align}
In view of these equations, (\ref{RT-left-pt}) and
(\ref{RT-right-pt}) coincide with (\ref{left-RT}) and
(\ref{right-RT}),  respectively. Furthermore, if we multiply the
second equation in (\ref{RT-left-pt}) and (\ref{RT-right-pt})
respectively by $A^l$ and $A^r$, and take the modulus square of the
resulting equations, we find that $|N_{l/r}|$ is a solution of
(\ref{eqn-pt}).

This completes our  demonstration of the equivalence of the
transfer-matrix approach to the solution of the scattering problem
for the nonlinear point interaction~(\ref{eqn-pt}) to the approach
we outline in Sec.~\ref{S3}. As seen from the above analysis, the
former has a simpler structure than the latter.

\section{Nonlinear double-$\delta$-function potential}
\label{S5}

Consider the scattering problem for the nonlinear interaction \cite{moya}:
    \be
    \begin{aligned}
    &\cF(x,\psi):=v(x,\psi)\psi(x),\\
    &v(x,\psi):=
    \sum_{\ell=1}^2
    f_\ell(|\psi(x)|)\,\delta(x-c_\ell),
    \end{aligned}
    \label{2pt}
    \ee
where $f_1$ and $f_2$ are real or complex-valued functions, and $c_1$
and $c_2$ are real parameters. $v(x,\psi)$ defines a
double-$\delta$-function potential. Because (\ref{2pt}) is of the
form (\ref{F=FF}) with
    \be
    \cF_\ell(x,\psi):=f_\ell(|\psi(x)|)\,\psi(x)\delta(x-c_\ell),
    \ee
$\cF_1(x,\psi)=0$ for $x>c:=(c_1+c_2)/2$, and $\cF_2(x,\psi)=0$ for
$x<c$, we can determine the transfer matrix $\bM$ of (\ref{2pt})
using the transfer matrix $\bM^{(\ell)}$ of $\cF_\ell(x,\psi)$ and
the composition rule (\ref{compose}).

$\bM^{(\ell)}(A_-,B_-)$ has the form given by (\ref{M-pt}) with $c$ and ${\fg}$ respectively replaced by $c_\ell$ and $i f_\ell(|e^{2ic\fK}A_-+B_-|)/2\fK$.
This together with  (\ref{compose}) and (\ref{AB-zero}) allow us to derive the following expressions for the entries of $\bM$.
    \bea
    M_{11}&=&1-\fg_1-\fg_2+(1-w^*) \fg_1\fg_2,
    \label{M11}\\
    M_{12}&=&e^{-2ic_1\fK}\left[-\fg_1-w^*\fg_2+(1-w^*)\fg_1\fg_2\right],
    \label{M12}\\
    M_{21}&=&e^{2ic_1\fK}\left[\fg_1+w \fg_2+(1-w)\fg_1\fg_2\right],
    \label{M21}\\
    M_{22}&=&1+\fg_1+\fg_2+(1-w) \fg_1\fg_2,
    \label{M22}
    \eea
where
    \begin{align}
    &\fg_\ell:=\frac{if_1(x_\ell)}{2\fK},
    \quad\quad w:=e^{2i(c_2-c_1)\fK},
    \label{ys}\\
    &x_1:=\left|e^{2ic_1\fK} A_-+B_-\right| ,
    \label{m1}\\
    &x_2:=\left|e^{2ic_2\fK} A_0+B_0\right|,
    \label{m1}\\
    &A_0:=(1-\fg_1)A_--e^{-2ic_1\fK}\fg_1B_-,
    \label{A-zero}\\
    &B_0:=e^{2ic_1\fK}\fg_1A_-+ (1+\fg_1)B_-.
    \label{A-zero}
    \end{align}
Again because $\det\bM^{(\ell)}=1$, we have $\det\bM=1$, which implies $\det\bM^{l}=1$. Therefore, to determine the reflection and transmission amplitudes of the double-$\delta$-function potential (\ref{2pt}) we need to compute $M_{21}^{l}$, $M_{22}^{l}$, $M_{12}^{r}$, and $M_{22}^{r}$.

According to (\ref{M-left-right}), $M_{21}^{l}$ and $M_{22}^{l}$ are respectively given by the right-hand side of (\ref{M21}) and (\ref{M22}) provided that we use $A_-=A^{l}$ and $B_-=A^{l} R^{l}$ to compute $\fg_\ell$. This gives
    \be
     \fg_\ell=\fg_\ell^{l}:=\frac{i f_\ell(x_\ell^l)}{2\fK},
     \label{g=g2}
     \ee
where
    \begin{align}
    x_1^l&:=\left|A^lC\right|,
    \label{x1=}\\
    C&:=1+e^{-2ic_1\fK}R^{l},
    \label{C=}\\
    x_2^l&:=\left|A^l\right|\left| [\fg_1^l(1-w)+1]C+w-1\right|.
    \label{x2}
    \end{align}
Substituting (\ref{g=g2}) in (\ref{M21}) and (\ref{M22}) and making use of (\ref{RT-left}), we have
    \bea
    R^{l}&=&-\frac{e^{2ic_1\fK}[\fg_1^{l}+w \fg_2^{l}+(1-w)\fg_1^{l} \fg_2^{l}]}{
    1+\fg_1^{l}+\fg_2^{l}+(1-w)\fg_1^{l} \fg_2^{l}},
    \label{2pt-R-left}\\
    T^{l}&=&\frac{1}{1+\fg_1^{l}+\fg_2^{l}+(1-w)\fg_1^{l} \fg_2^{l}}.
    \label{2pt-T-left}
    \eea

Next, we solve (\ref{2pt-T-left}) for $\fg^l_2$ to obtain
    \be
    \fg_2^l=\frac{1-(1+\fg_1^l)T^l}{T^l[(1-w)\fg_1^l+1]},
    \label{eqz1}
    \ee
and use (\ref{C=}), (\ref{2pt-R-left}), and (\ref{2pt-T-left}), to show that
    \be
    \frac{C}{T^l}=(1-w)\fg_2^l+1.
    \label{eqz2}
    \ee
Inserting (\ref{eqz1}) in (\ref{eqz2}) gives
    \be
    C=\frac{1-w+w T^l}{(1-w)\fg_1^l+1}.
    \nn
    \ee
If we substitute this equation in (\ref{x2}), we arrive at the remarkable relation: $x_2^l=\left|A^lT^l\right|=|N_l|$, which in view of (\ref{g=g2}) implies:
    \begin{align}
    &\fg_2^l=g_2(\left|N_l\right|),
    &&g_2(x):=\frac{i}{2\fK}f_2(x).
    \label{g2-left}
    \end{align}

Furthermore, we can use (\ref{g=g2}), (\ref{x1=}), (\ref{eqz2}), and (\ref{g2-left}) to show that
    \be
    \fg_1^l=g_1(\left|N_l\right|),
    \label{g1-left}
    \ee
where
    \bea
    g_1(x)&:=&\frac{i}{2\fK}f_1\left(x\left|(1-w)g_2(x)+1\right|\right)\nn\\
    &=&\frac{i}{2\fK}f_1\big(x\left| i(1-w)f_2(x)/2\fK+1\right|\big).~~~
    \label{g1-left-def}
    \eea
If we multiply both sides of (\ref{2pt-T-left}) by $A^l$ and equate their modulus, we find that $|N_l|$ is a real and nonnegative solution of the following real equation.
    \be
    x\left|(1-w)g_1(x)g_2(x)+g_1(x)+g_2(x)+1\right|-\left|A^l\right|=0.
    \label{2pt-main-eq}
    \ee
Given such a solution, we can identify it with $|N_l|$ and use it in (\ref{g2-left}) and (\ref{g1-left}) to determine $\fg_\ell^l$. Plugging the result in (\ref{2pt-R-left}) and (\ref{2pt-T-left}), we obtain $R^l$ and $T^l$.

The calculation of $R^r$ and $T^r$ is similar. First, we set $A_-=0$ and $B_-=A^rT^r$ in (\ref{ys}) -- (\ref{A-zero}) to establish
    \be
    \begin{aligned}
    &\fg_1=\fg_1^r:=\frac{i}{2\fK}f_1(|N_r|),\\
    &\fg_2=\fg_2^r:=\frac{i}{2\fK}f_2\big(|N_r|\left|i(1-w)f_1(|N_r|)/2\fK+1\right|\big).
    \end{aligned}
    \label{ys-right}
    \ee
Substituting these equations in (\ref{M12}) and (\ref{M22}) and making use of (\ref{RT-right}), we find
    \bea
    R^r&=&-\frac{e^{-2ic_1\fK}[\fg_1^r+w^*\fg_2^r-(1-w^*)\fg^r_1\fg^r_2]}{
    1+\fg_1^{r}+\fg_2^{r}+(1-w)\fg_1^{r} \fg_2^{r}},
    \label{2pt-R-right}\\
    T^r&=&\frac{1}{1+\fg_1^{r}+\fg_2^{r}+(1-w)\fg_1^{r} \fg_2^{r}}.
    \label{2pt-T-right}
    \eea
Next, we multiply both sides of (\ref{2pt-T-right}), take their modulus, and recall that $N_r:=A^rT^r$. This show that $|N_r|$ satisfies (\ref{2pt-main-eq}), if we redefine $g_\ell(x)$ according to:
    \be
    \begin{aligned}
    g_1(x)&:=\frac{i}{2\fK}f_1(x), \\
    g_2(x)&:=\frac{i}{2\fK}f_2\big(x\left|i(1-w)f_1(x)/2\fK+1\right|\big).
    \end{aligned}
    \label{g-x-r}
    \ee
Comparing these  relations respectively with (\ref{g1-left}) and
(\ref{g2-left}), we see that they coincide, if we swap the roles of
the functions $f_1(x)$ and $f_2(x)$. Again given a real and positive
solution of (\ref{2pt-main-eq}) with $g_\ell(x)$ given by
(\ref{g-x-r}) we can determine $|N_r|$ and consequently
$\fg_\ell^r$, $R^r$, and $T^r$.

As our treatment of the nonlinear double-$\delta$-function potential shows, the application of the nonlinear transfer matrix in computing $R^r$ and $T^r$ is simpler than that of $R^l$ and $T^l$. We could have avoided the nontrivial analysis we employed in our derivation of $R^l$ and $T^l$ by trying to use the transformation properties of $R^{l/r}$ and $T^{l/r}$ under a space reflection \cite{jpa-2014c,bookchapter}. This allows for reducing the problem of finding $R^l$ and $T^l$ for a given nonlinear scattering interaction to the determination of $R^r$ and $T^r$ for its reflection in space, i.e., the parity transformed interaction. For the double-$\delta$-function potential~(\ref{2pt}), this argument implies that we can obtain $R^l$ and $T^l$ by changing $A^r$ to $A^l$ and swapping $(c_1,f_1)$ and $(c_2,f_2)$ in the expression for $R^r$ and $T^r$, respectively. This is in complete agreement with our direct calculation of $R^{l/r}$ and $T^{l/r}$.

Our results show that the scattering problem scattering problem for nonlinear double-$\delta$-function potential reduces to the solution of (\ref{2pt-main-eq}). Because this equation does not admit a closed-form analytic solution, it is not possible to determine the explicit dependence of  $R^{l/r}$ and $T^{l/r}$ on $\fK$ and $|A^{l/r}|$. We can however provide a graphical demonstration of the behavior of  $R^{l/r}$ and $T^{l/r}$ for particular choices of $f_\ell(x)$ and various ranges of values of $\fK$ and $|A^{l/r}|$.

\begin{figure*}
    \begin{center}
    \includegraphics[scale=.43]{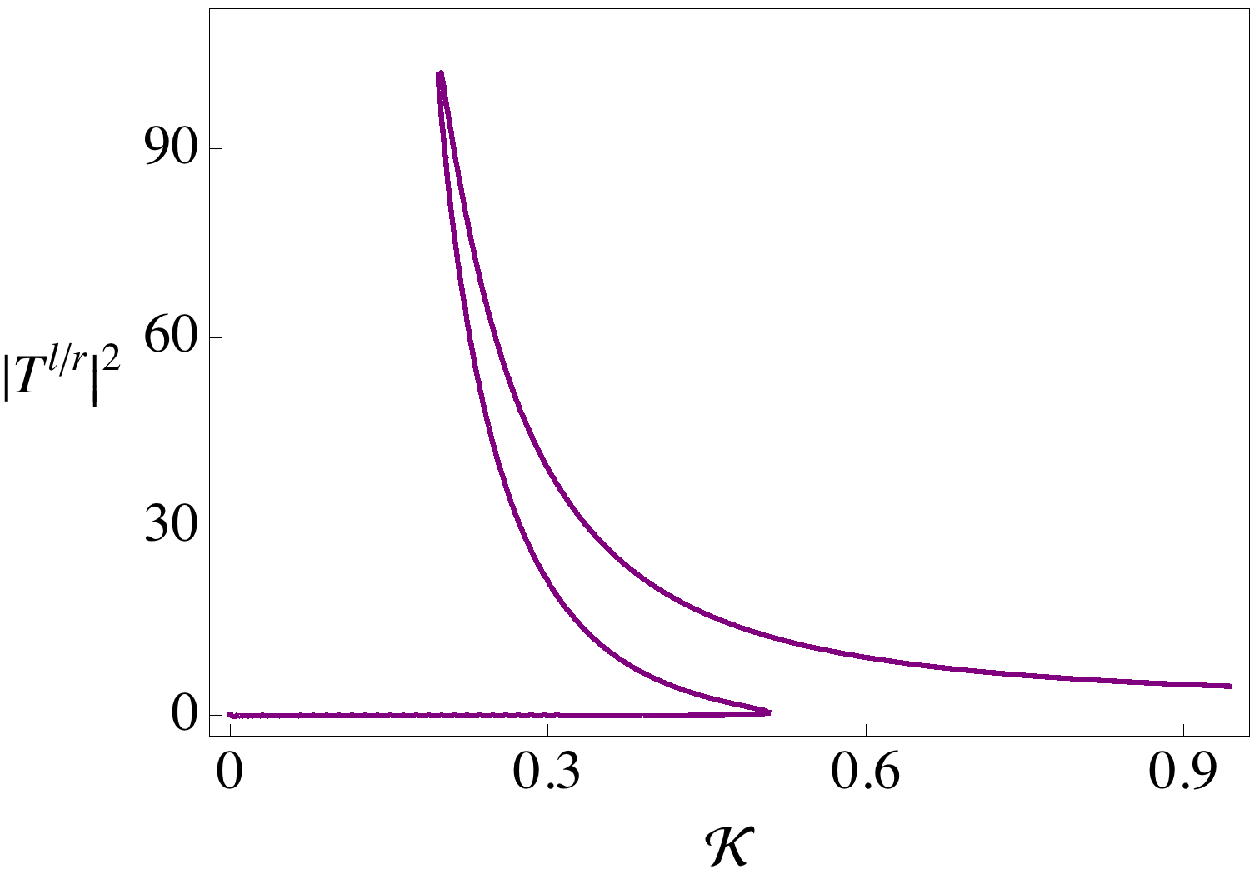}~~
    \includegraphics[scale=.43]{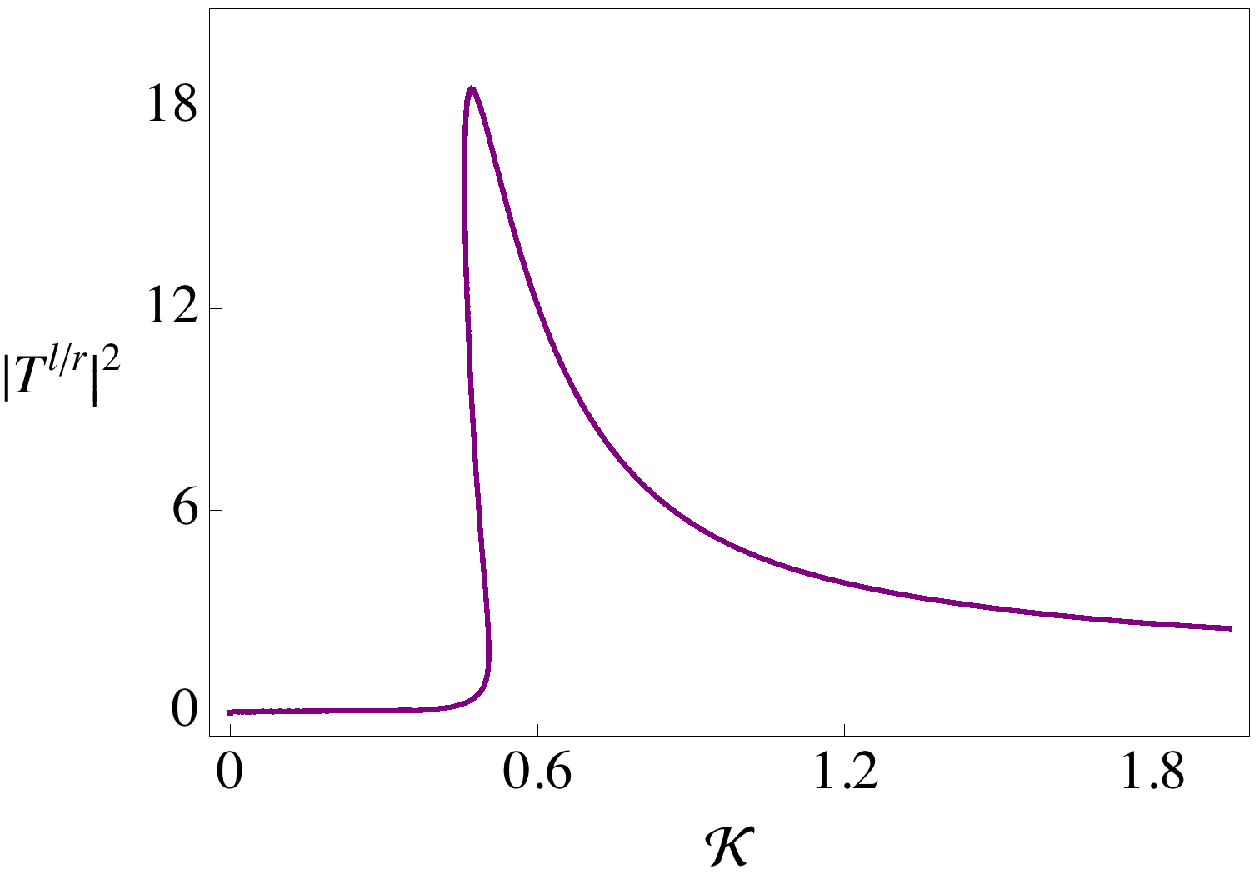}~~
    \includegraphics[scale=.43]{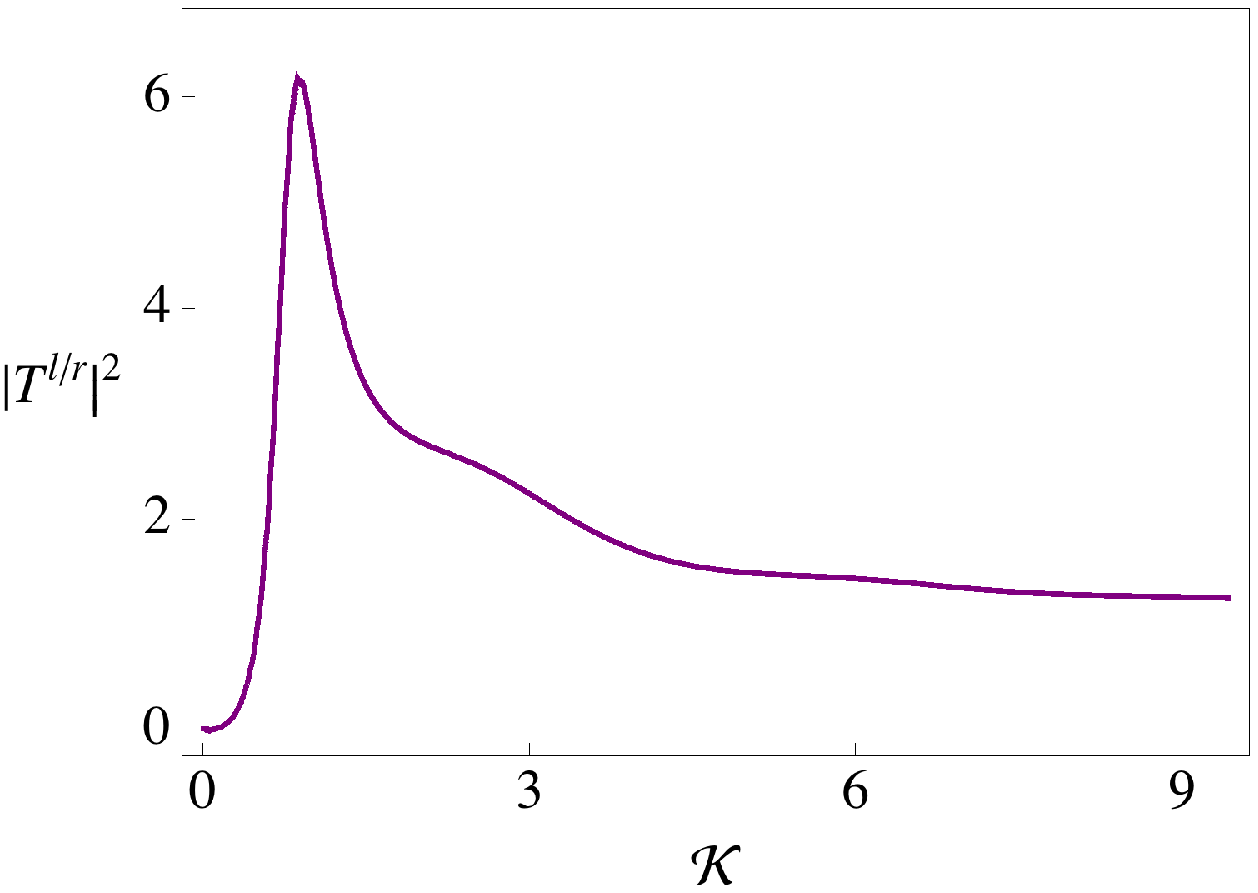}\\
    \includegraphics[scale=.43]{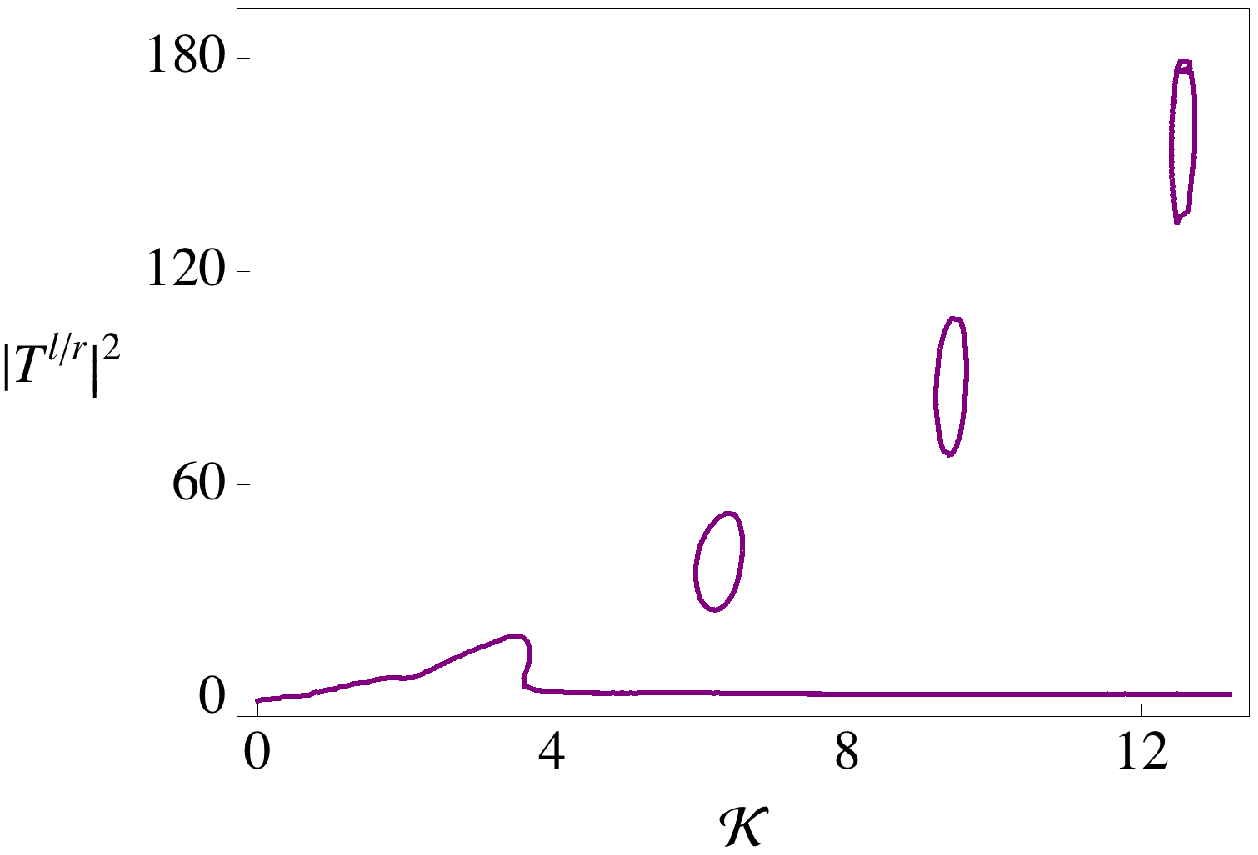}~~
    \includegraphics[scale=.43]{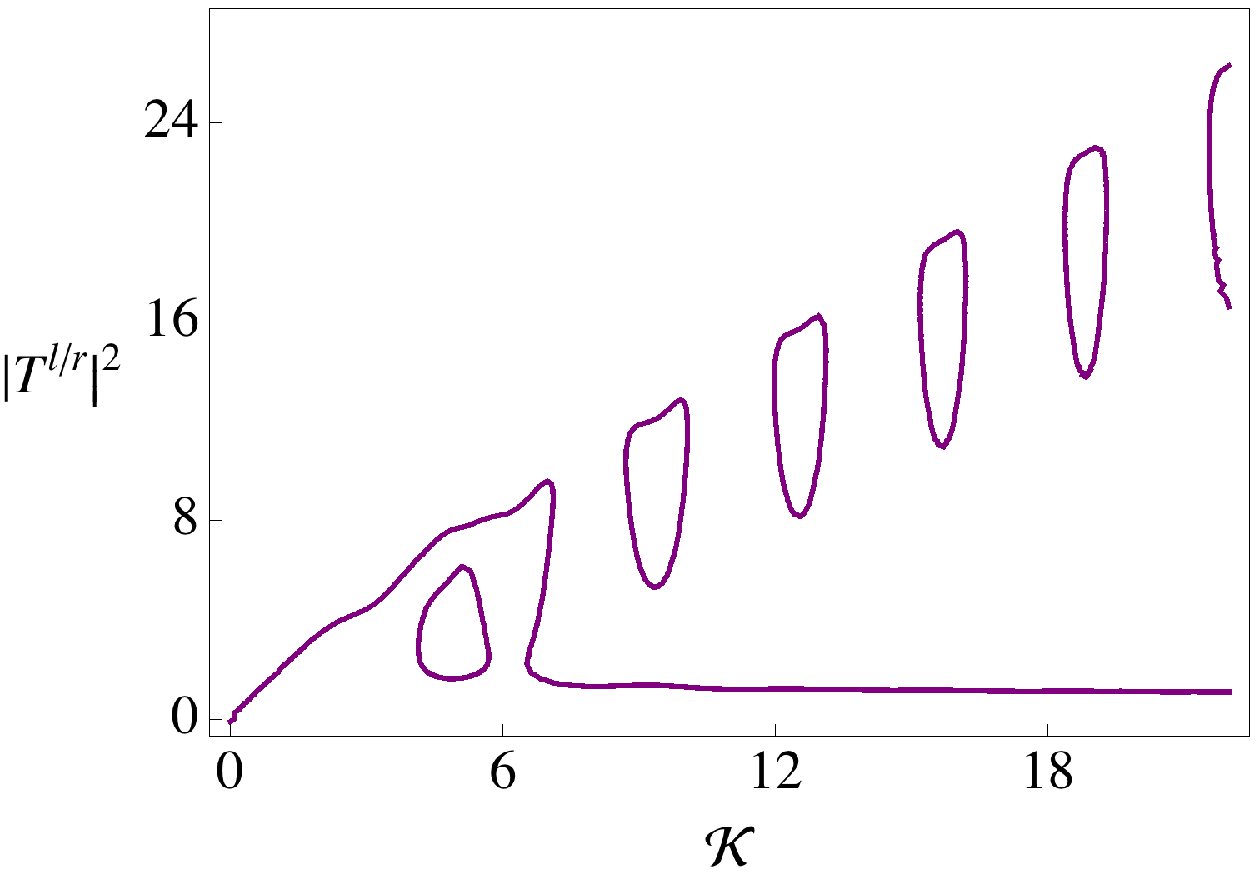}~~
    \includegraphics[scale=.43]{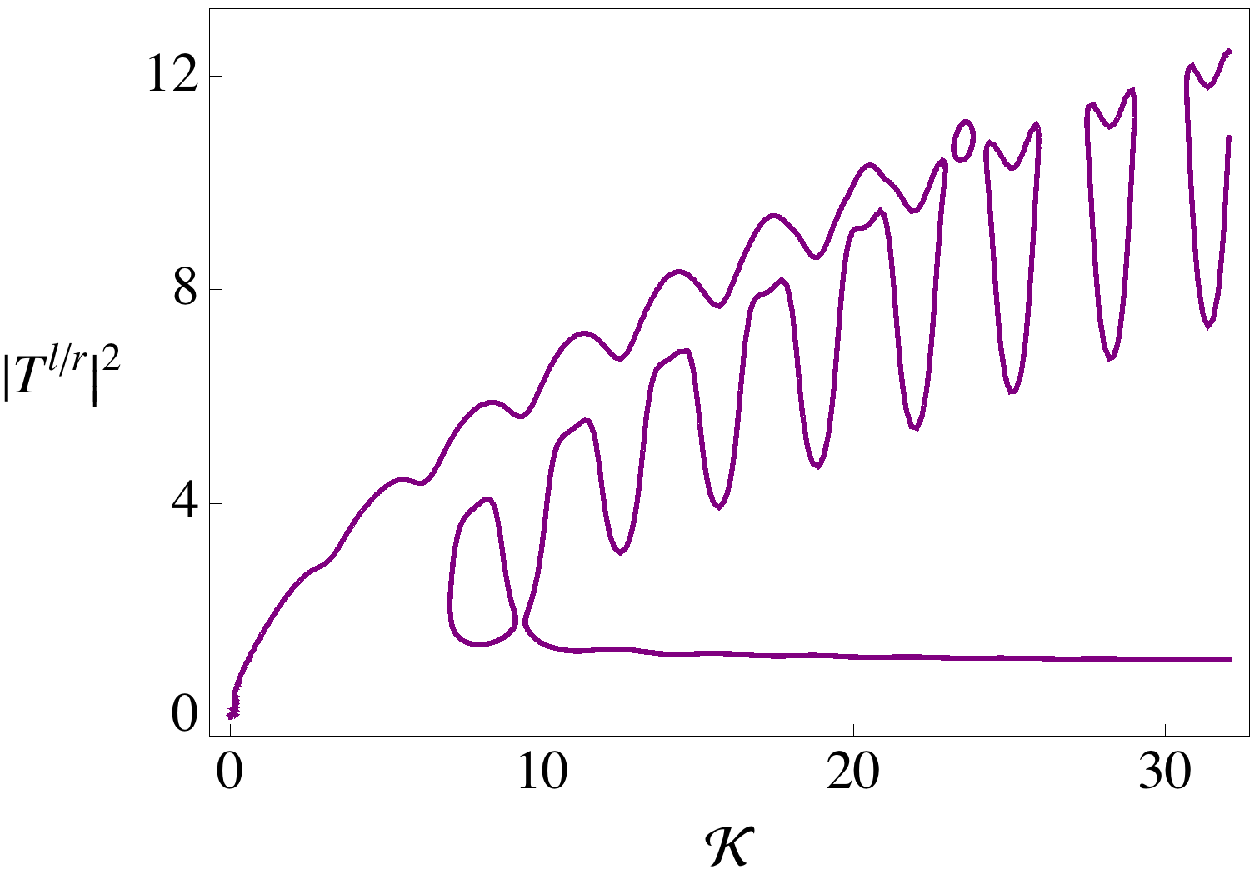}
    \caption{Plots of $|T^{l/r}|^2$ for the $\cP$-symmetric
    double-$\delta$-function potential (\ref{2p-power}) with
    $\fz_\ell=i$, $\nu_\ell= -0.7$ (top left), $\nu_\ell=-0.5$
    (top middle), $\nu_\ell=0$ (top right), $\nu_\ell=1$ (bottom left),
    $\nu_\ell=2$ (bottom middle), $\nu_\ell=3$ (bottom right), and
    $|A^{l/r}|=1$. The values of $|T^{l/r}|$ can exceed 1 because
    the imaginary part of the coupling constants is positive.}
    \label{fig2}
    \end{center}
    \end{figure*}

\begin{figure*}
   \begin{center}
   \includegraphics[scale=.43]{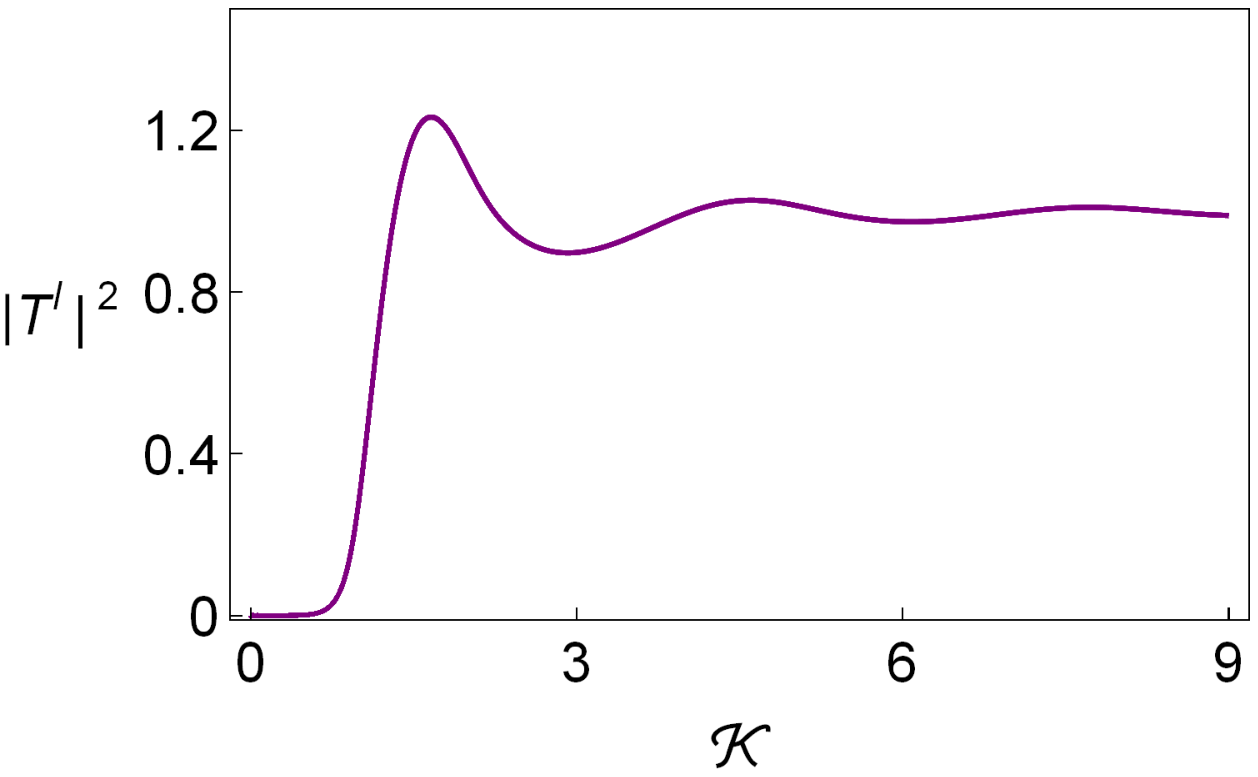}~~
   \includegraphics[scale=.43]{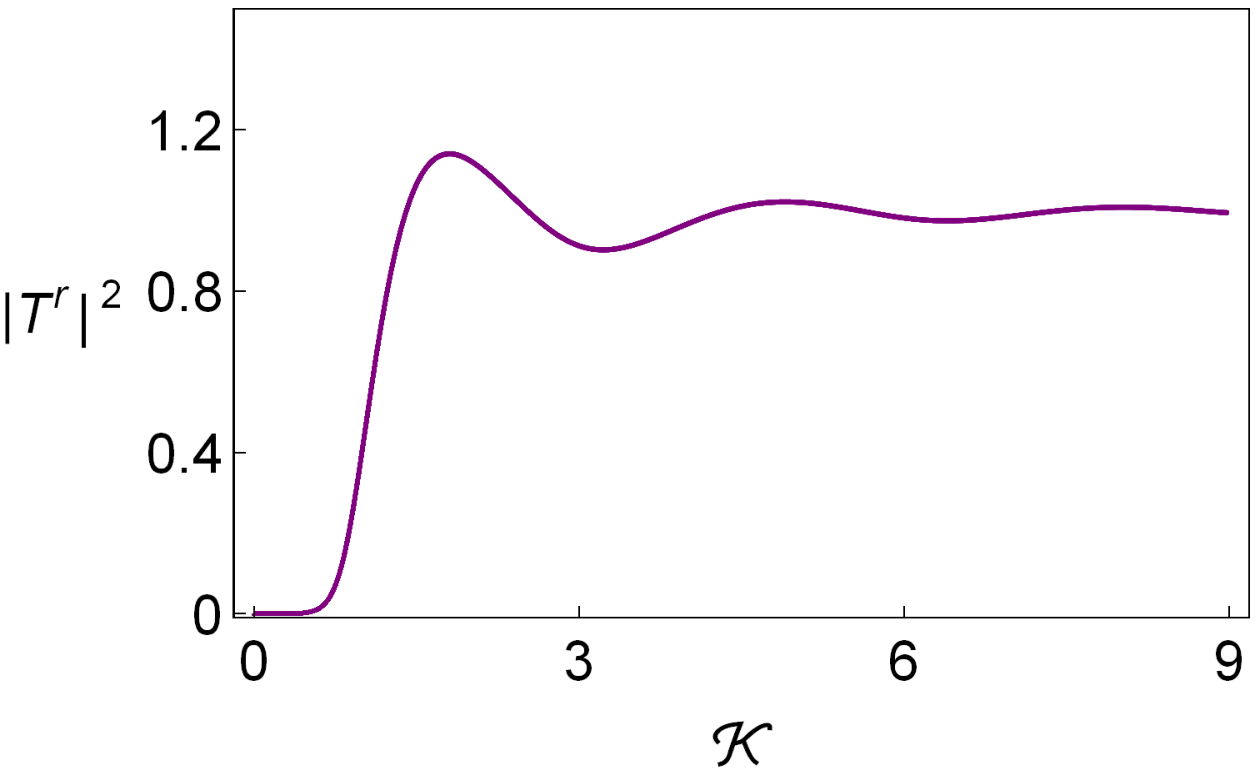}~~
   \includegraphics[scale=.43]{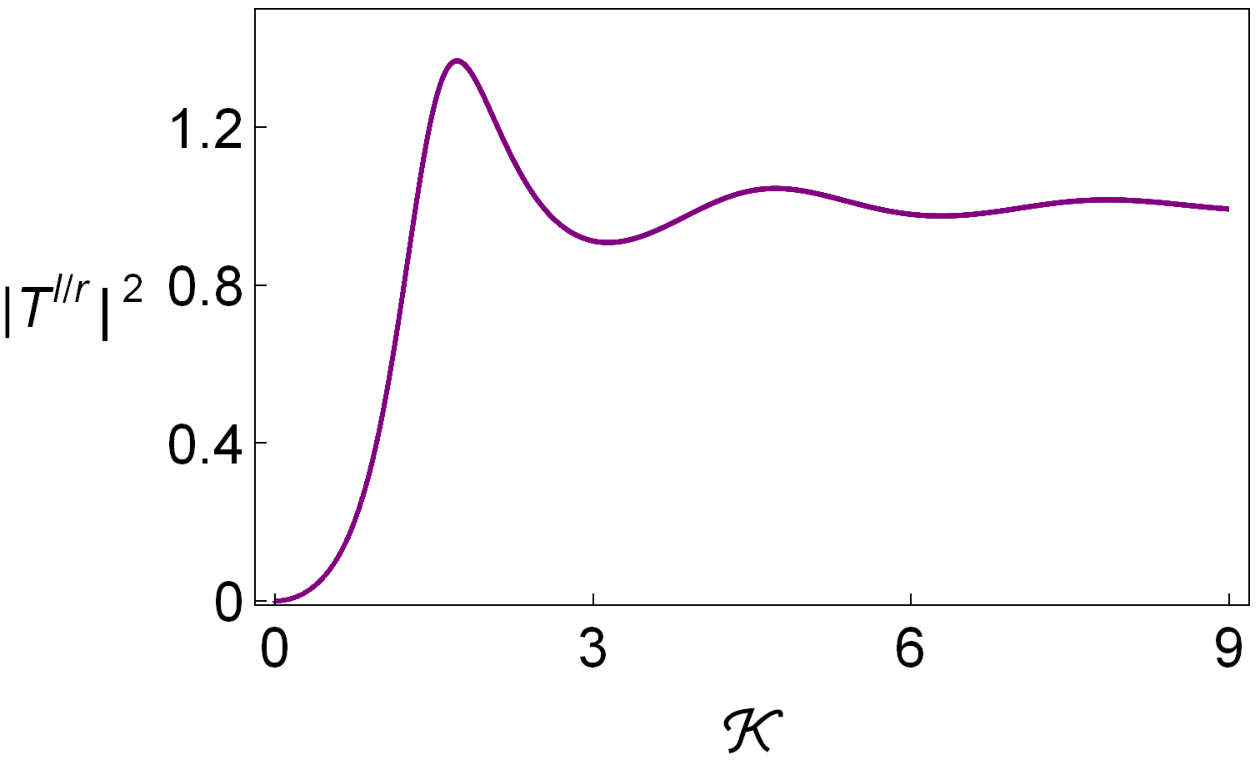}\\[6pt]
   \includegraphics[scale=.32]{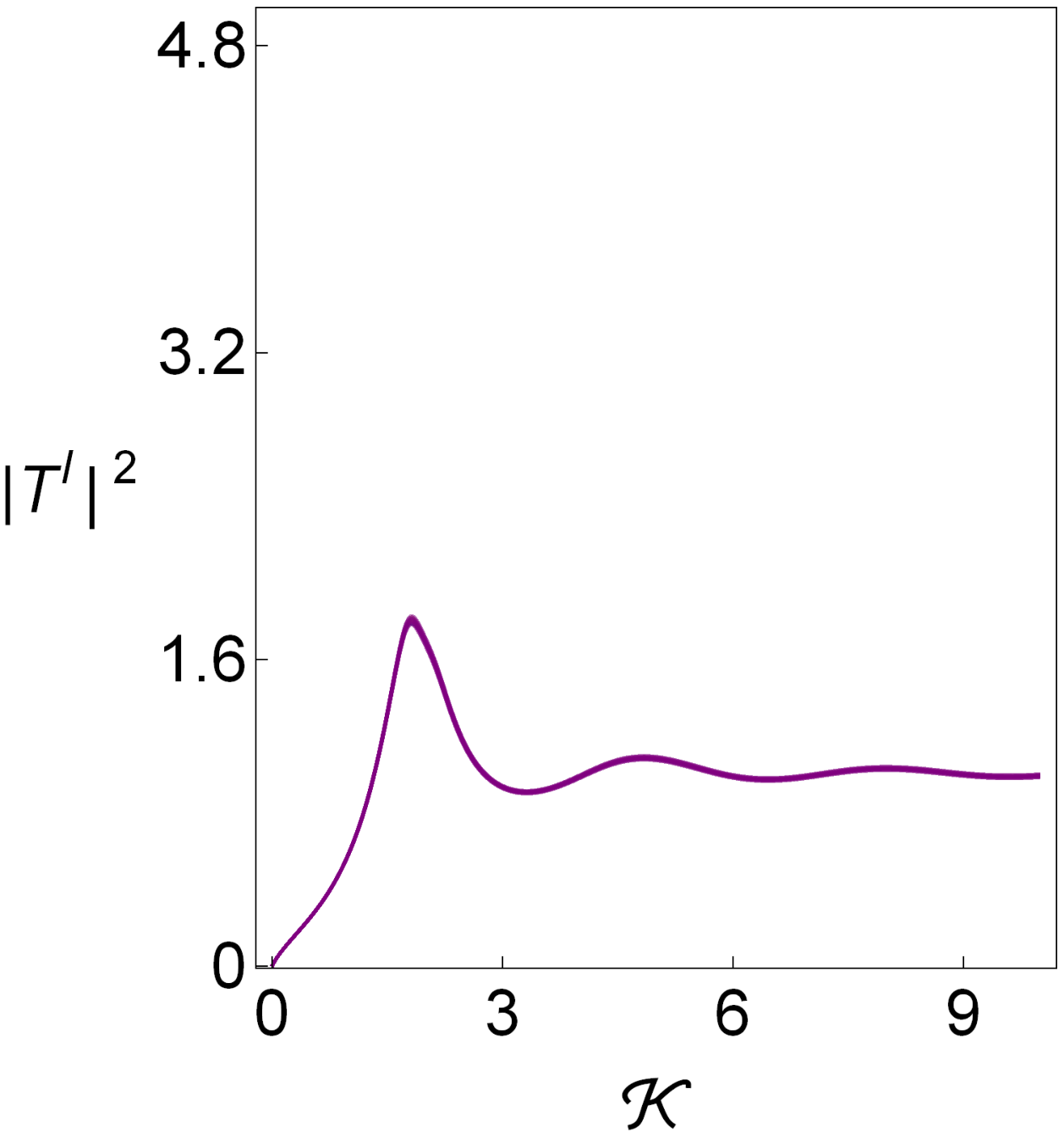}~~
   \includegraphics[scale=.32]{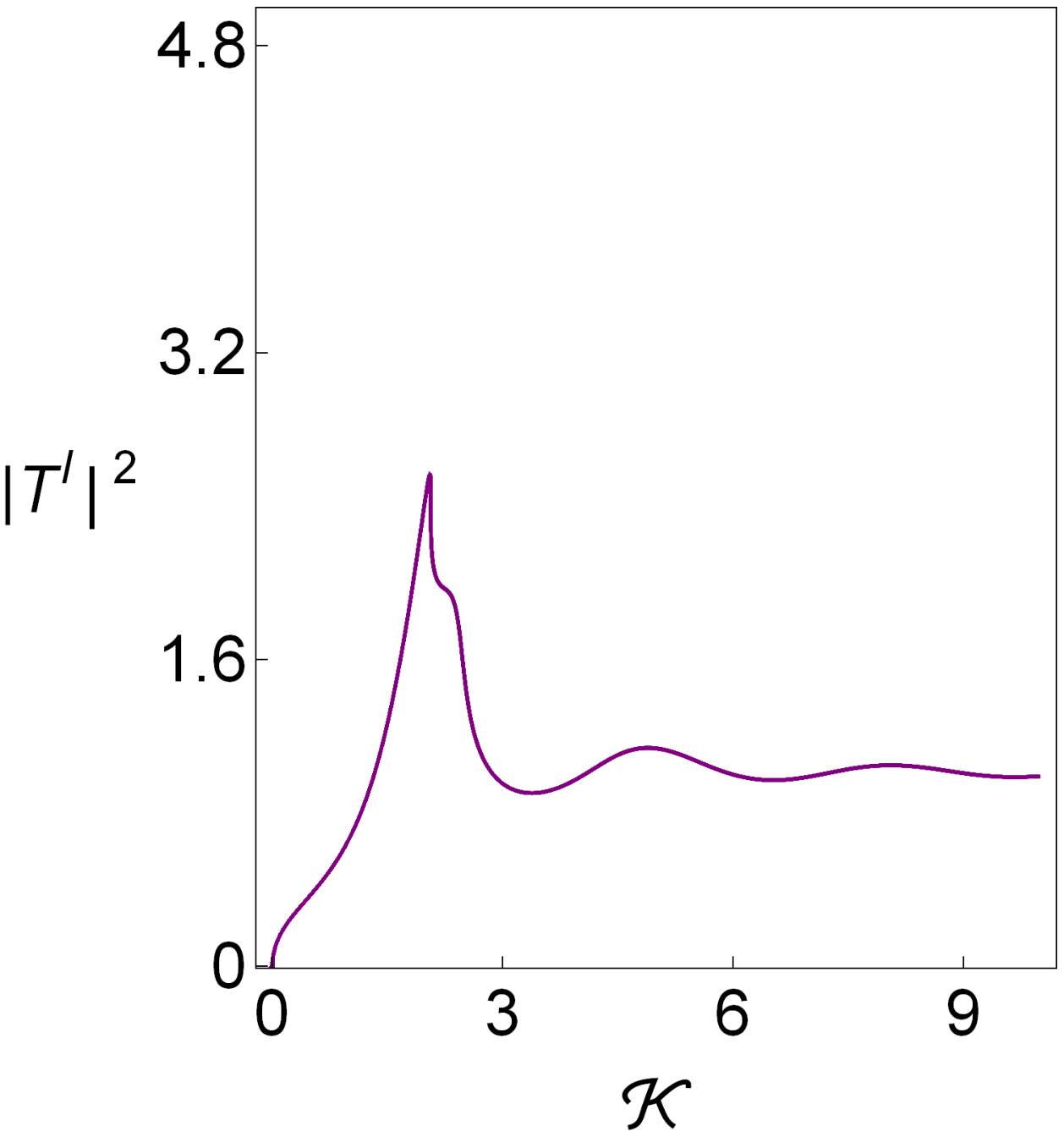}~~
   \includegraphics[scale=.32]{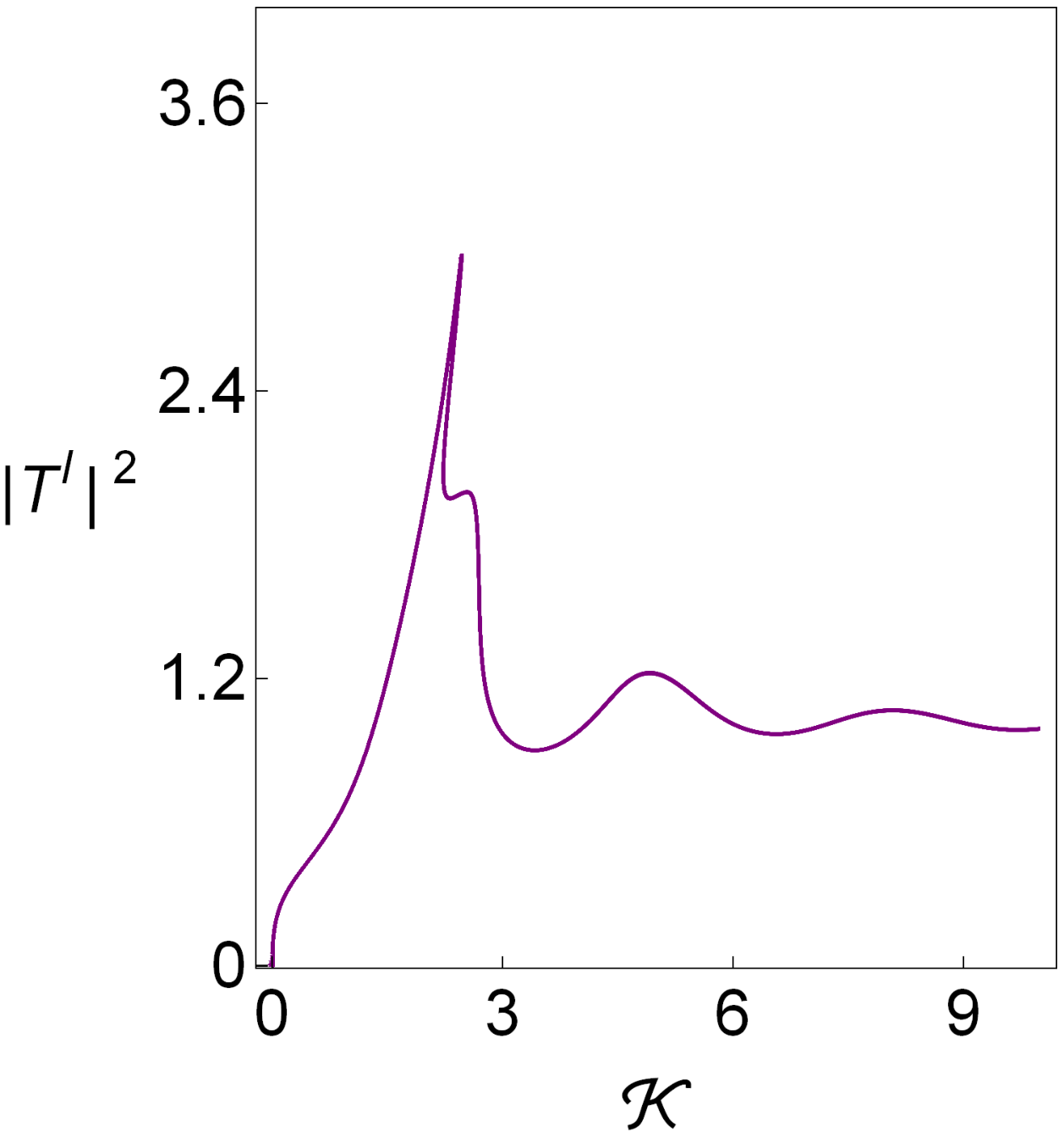}~~
   \includegraphics[scale=.32]{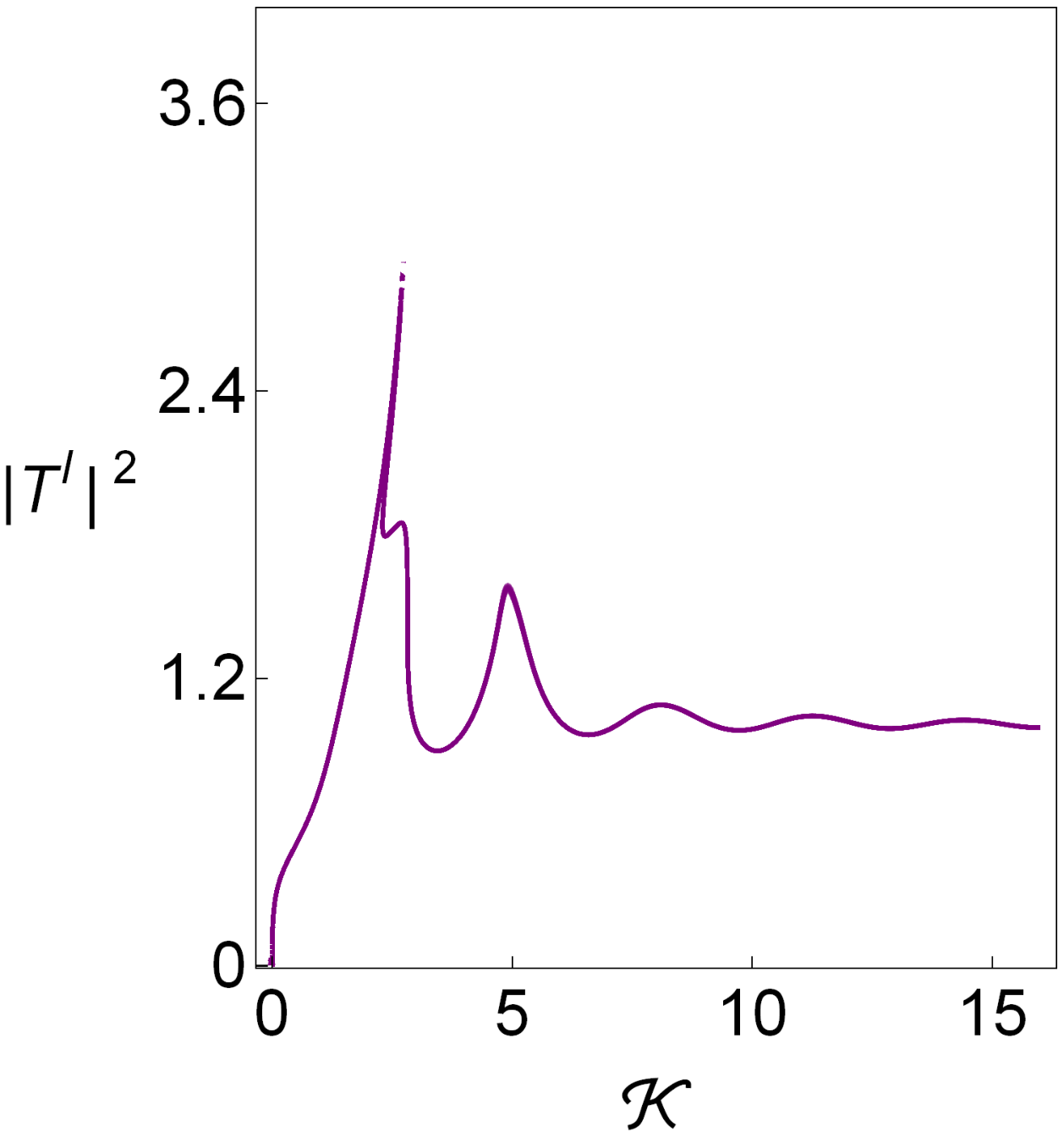}\\[6pt]
   \includegraphics[scale=.32]{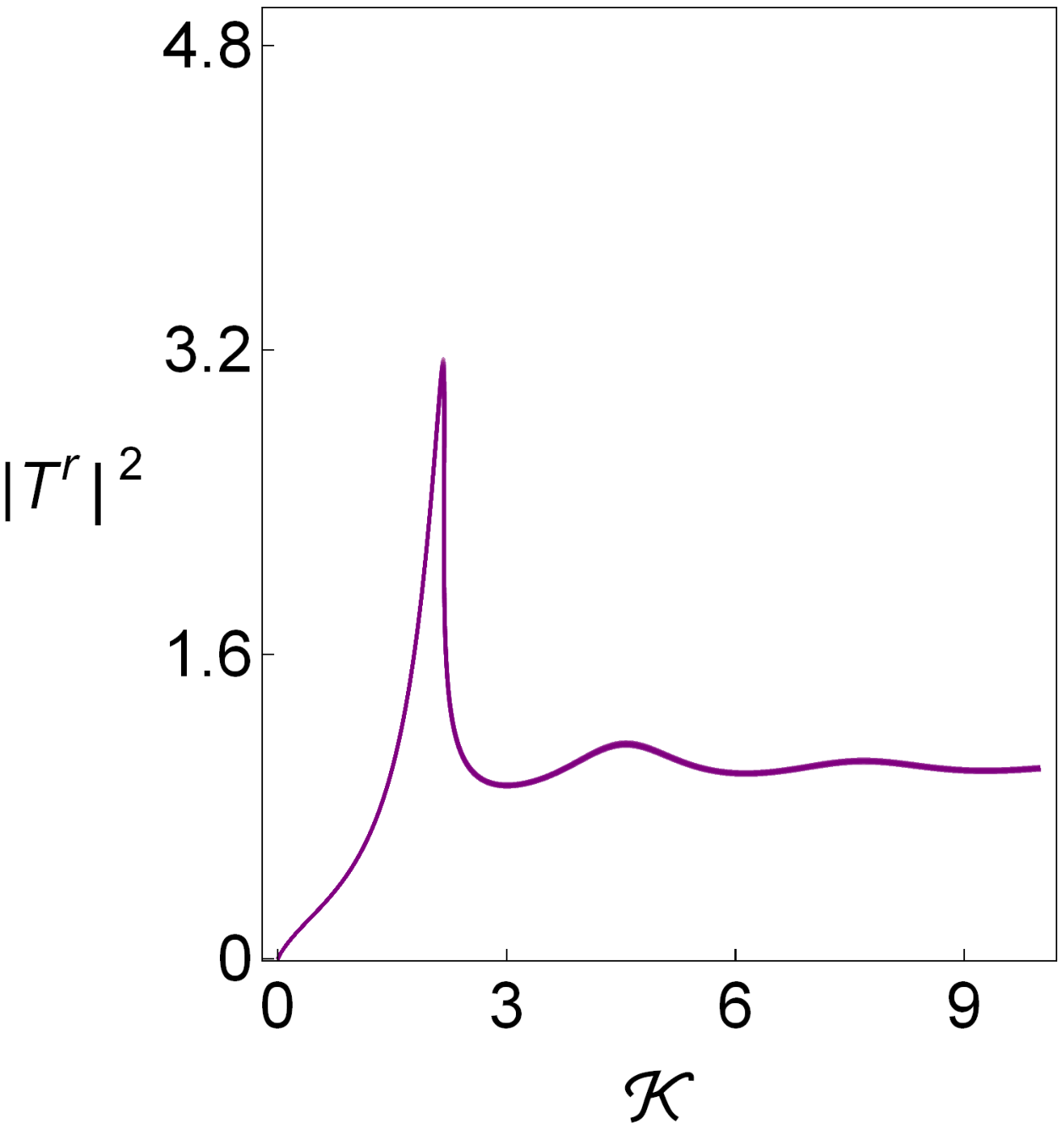}~~
   \includegraphics[scale=.32]{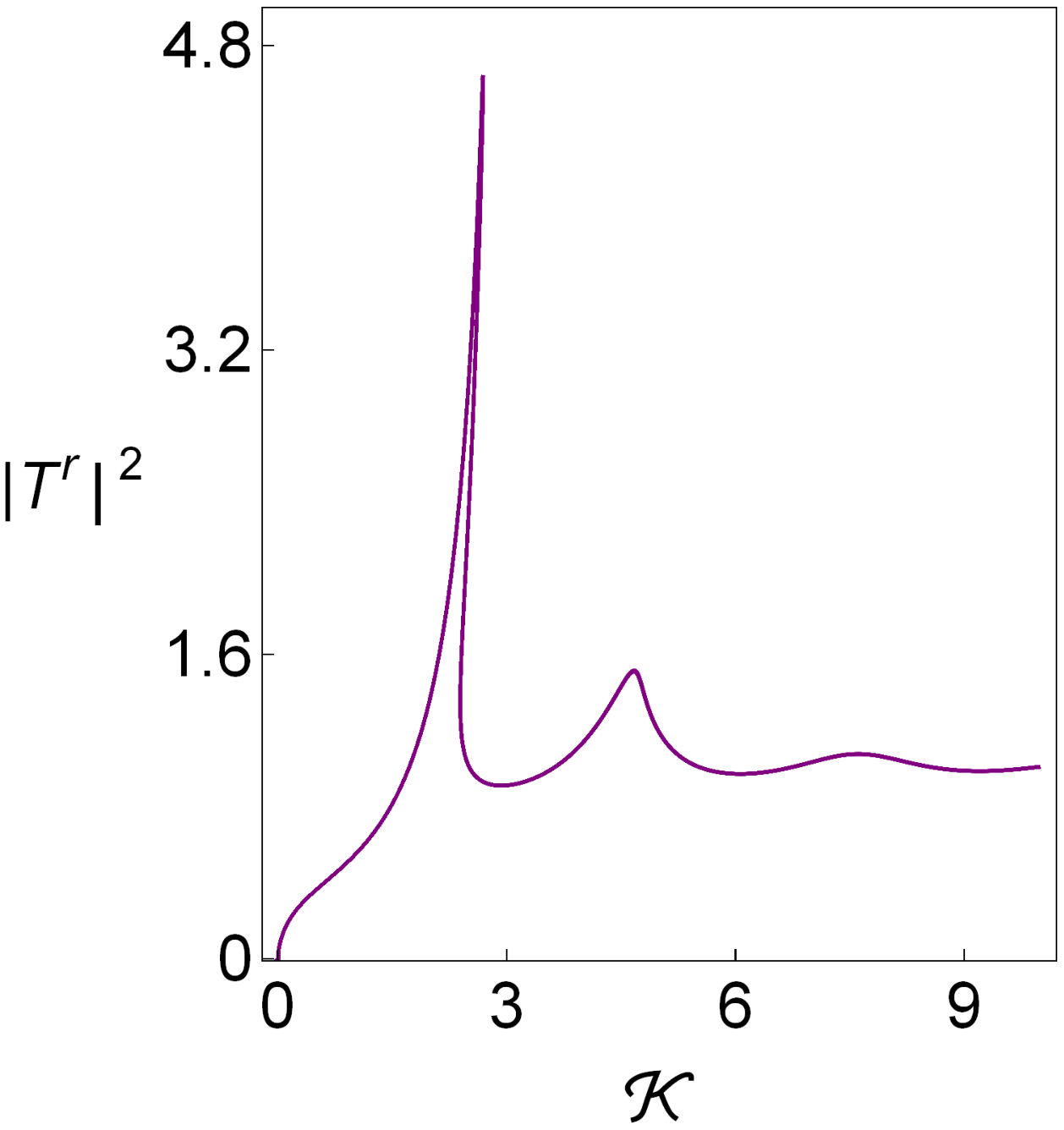}~~
   \includegraphics[scale=.32]{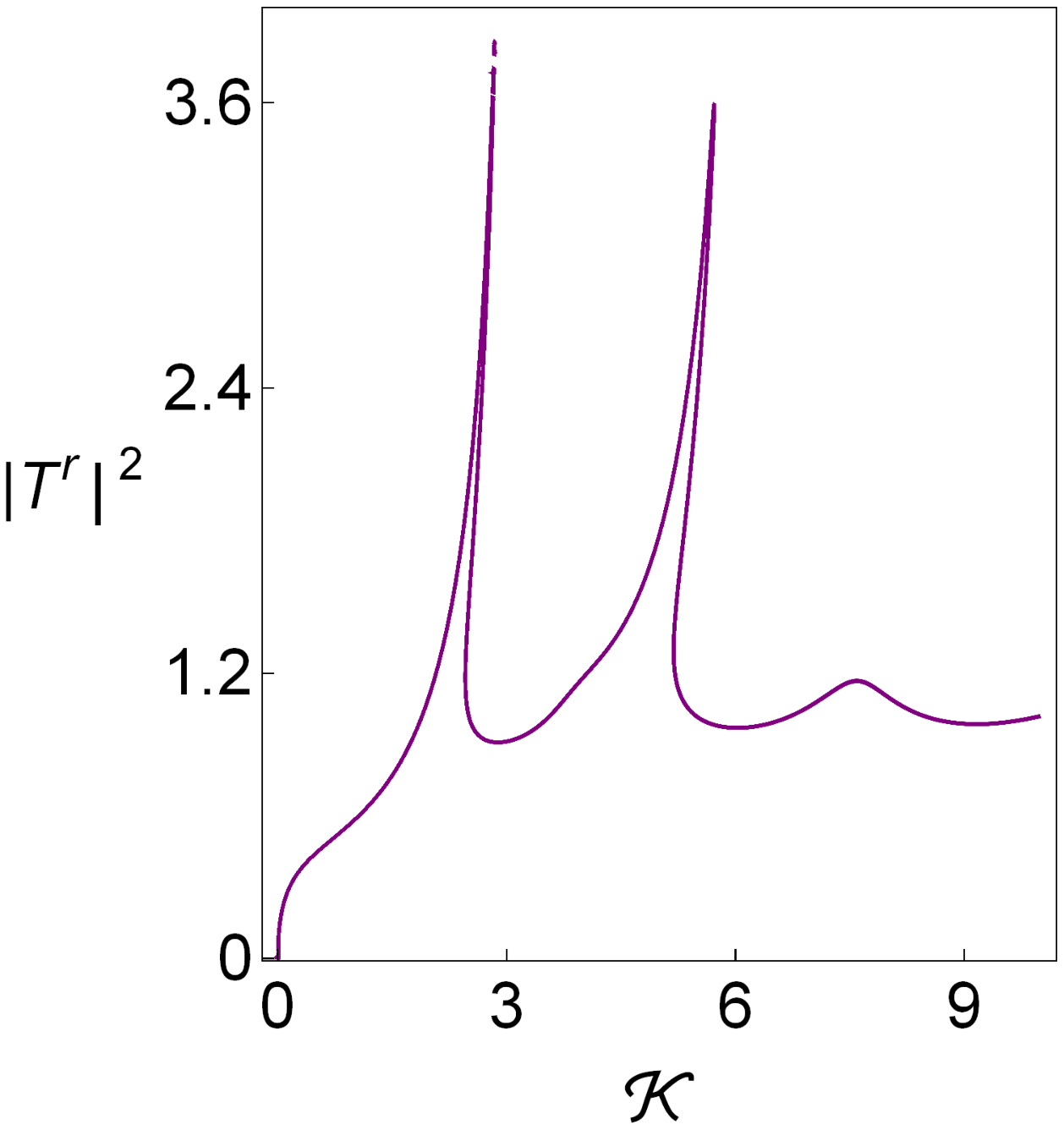}~~
   \includegraphics[scale=.32]{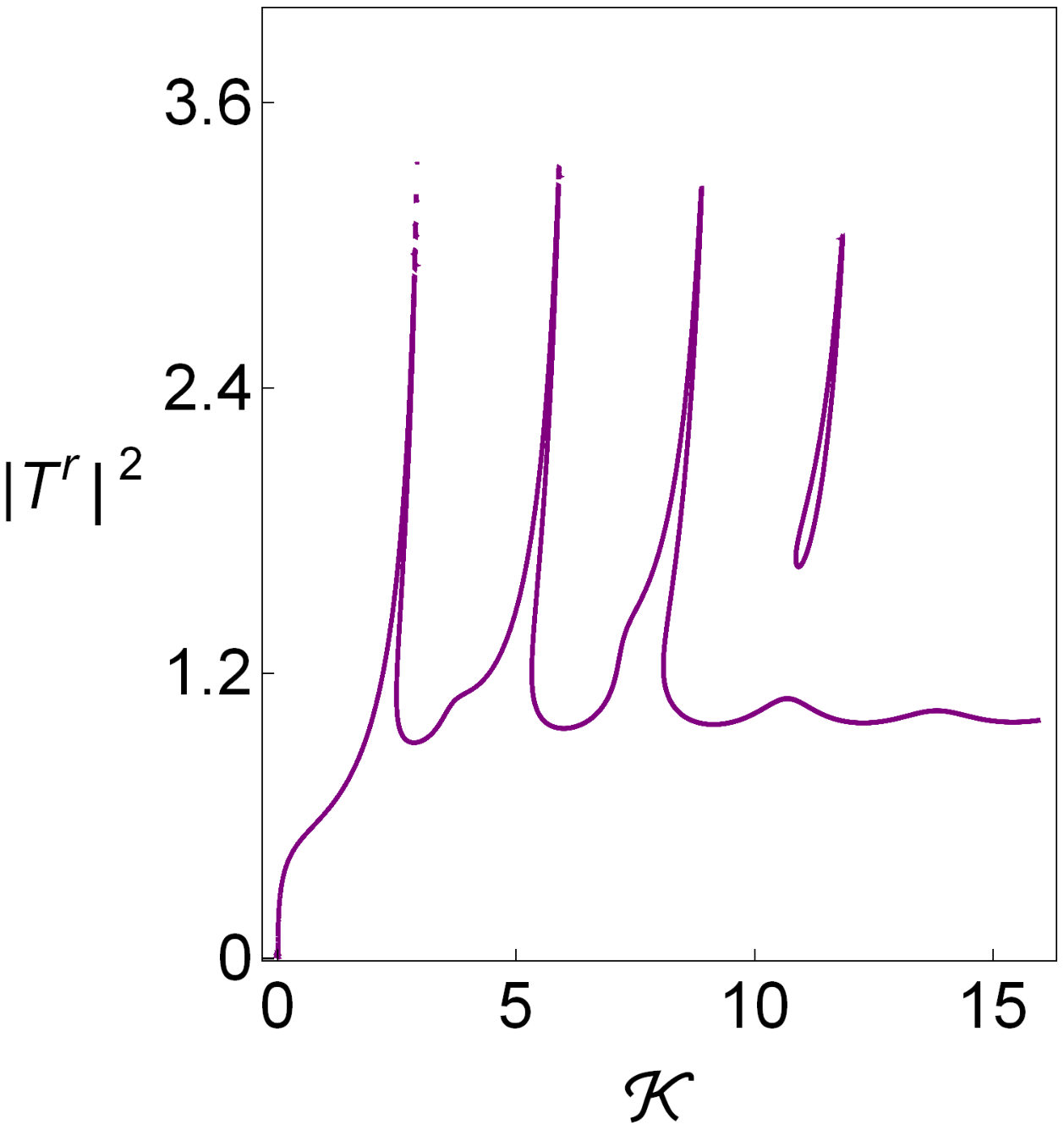}
   \caption{Plots of $|T^{l/r}|^2$ for the $\cP\cT$-symmetric
   double-$\delta$-function potential (\ref{2p-power}) with
   $\fz_1=\fz_2^*=1-i$, $\nu_\ell= -0.5$ (top left and top middle graphs),
   $\nu_\ell=0$ (top right graph), and $\nu_\ell=1,2,3,4$ (The middle and
   bottom are respectively the graphs of $T^l$ and $T^r$.) We have taken
   $|A^{l/r}|=1$. For $\nu_\ell\neq 0$, the difference between the graphs
   of $|T^l|^2$ and $|T^r|^2$ clearly demonstrates the violation of
   reciprocity in transmission.}
   \label{fig3}
   \end{center}
   \end{figure*}

\begin{figure*}
   \begin{center}
   \includegraphics[scale=.5]{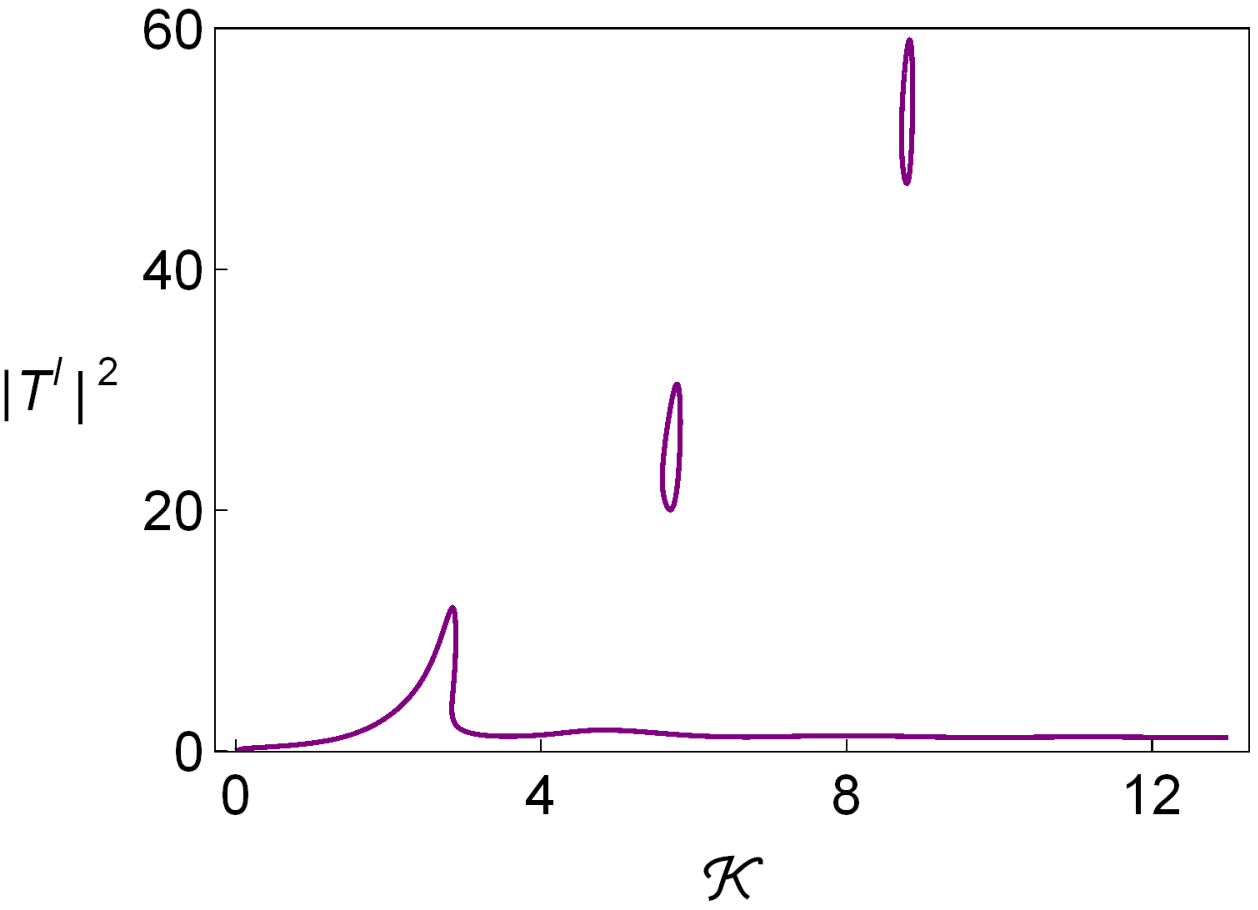}~~~~
   \includegraphics[scale=.5]{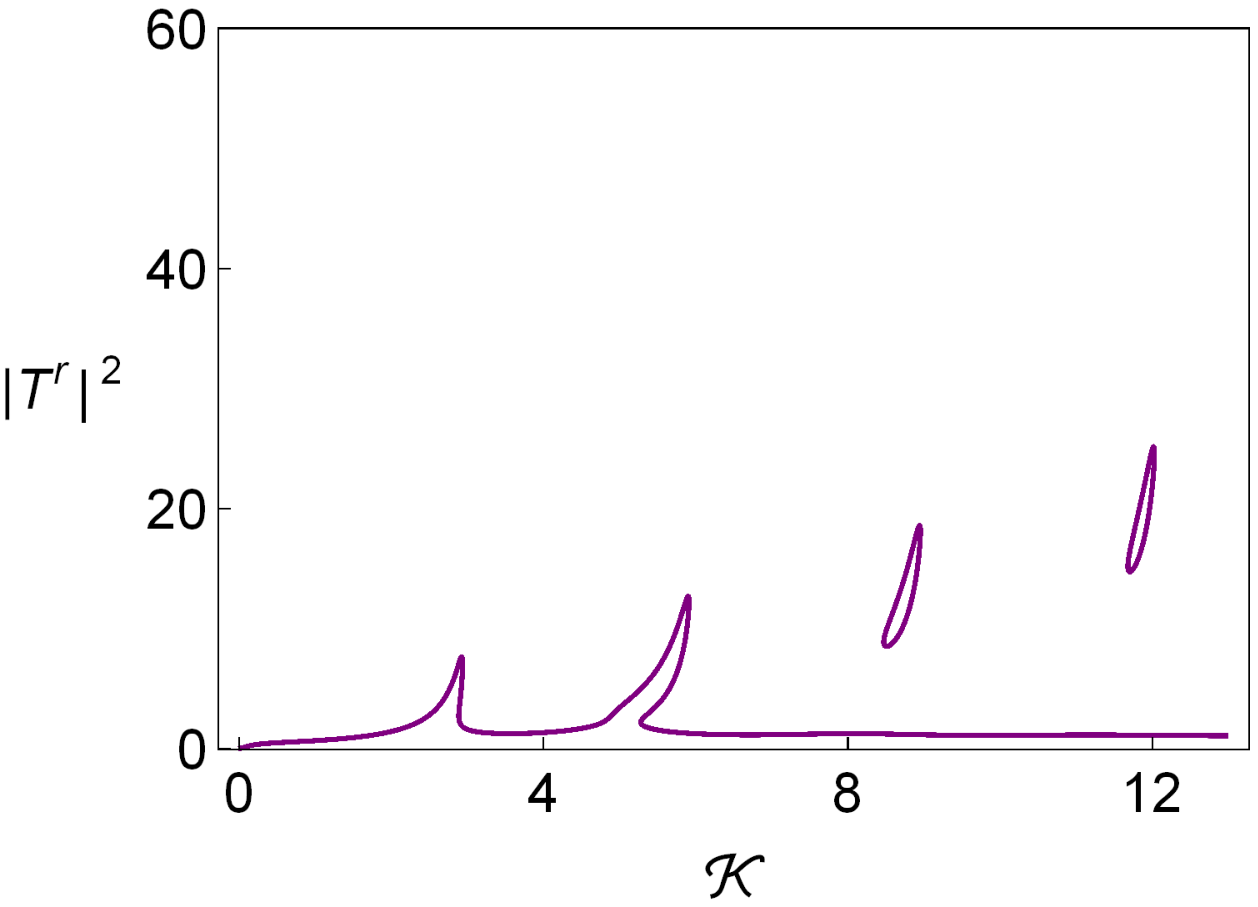}
   \caption{Plots of $|T^{l/r}|^2$ for a complex double-$\delta$-function potential (\ref{2p-power}) with $\fz_2=-2\fz_1=1+2i$, $\nu_1=2\nu_2=2$, and $|A^{l/r}|=1$. The difference between the graphs for $|T^l|^2$ and  $|T^r|^2$ demonstrates the violation of reciprocity in transmission.}
   \label{fig4}
   \end{center}
   \end{figure*}

Without  much effort we can use Mathematica to produce plots of
$|T^{l/r}|^2$ from (\ref{2pt-main-eq}). Figs.~\ref{fig2}--
\ref{fig4} give the graphs of $|T^{l/r}|^2$ as functions of $\fK$
for $|A^{l/r}|=1$, $c_\ell:=(-1)^\ell/2$ and $f_\ell(x):=\fz_\ell\,
x^{\nu_\ell}$ which determine the double-$\delta$-function
potential:
    \be
    v(x,\psi)=\fz_1\,|\psi(x)|^{\nu_{1}}\delta(x+\mbox{$\frac{1}{2}$})+
    \fz_2\,|\psi(x)|^{\nu_{2}}\delta(x-\mbox{$\frac{1}{2}$}),
    \label{2p-power}
    \ee
and the following choices for $\fz_\ell$, $\nu_\ell$.
    \begin{align}
    & \nu_\ell=-0.7, -0.5, 0, 1, 2, 3 &&\fz_\ell=i,
    \label{P-inv-real}\\
    & \nu_\ell=-0.5, 0, 1, 2, 3, 4 &&\fz_1=\fz_2^*=1- i,
    \label{PT-inv-real}\\
    & \nu_1=2\nu_2=2, &&\fz_2=-2\fz_1=1+2i.
    \label{gen}
    \end{align}
These correspond to certain $\cP$- and $\cP\cT$-symmetric
\cite{konotop-review} nonlinear double-$\delta$-function potentials,
as well as a complex double-$\delta$-function potential without such
symmetries. The graphs show rich transmission bistability and
multistability features particularly due to the fact that at least
one of the coupling constants has a positive imaginary part. In
optical realizations this corresponds to the presence of a gain
component that is responsible for the amplification of the wave and
leads to values of $|T^{l/r}|^2$ that exceed unity. For a detailed
examination of the transmission properties of a nonlinear
double-$\delta$-function potential with real coupling constants, see
\cite{moya}.



\section{Summary and conclusions}
\label{S6}

In this article we have  outlined a systematic treatment of
nonlinear scattering systems, which is based on the use of Jost
functions, and introduced a nonlinear analog of the transfer matrix
of standard linear scattering theory. Unlike its linear analog the
nonlinear transfer matrix is not uniquely determined by wave
equation describing the scattering phenomenon. This  does not
however obstruct its effectiveness in solving nonlinear scattering
problems. This is because we can characterize the ambiguity in the
definition of the nonlinear transfer matrix in terms of a pair of
functions that turn out not to enter the expressions for the
reflection and transmission amplitudes. Therefore, we can freely
choose these functions and use the resulting nonlinear transfer
matrix to address scattering problems.

The formulas giving the  reflection and transmission amplitudes in
terms of the entries of the nonlinear transfer matrix are almost
identical to their linear analogs. This allows for a simple
characterization of interesting scattering features such as
directional reflectionlessness and transparency, nonreciprocal
transmission, as well as coherent perfect emission and absorption of
waves which respectively correspond to the presence of (nonlinear)
spectral singularities and their time-reversal.

An important aspect of the  formulation of nonlinear scattering
problems using the nonlinear transfer matrix is its composition
property. This is a nonlinear generalization of the well-known
composition property of the linear transfer matrix. It makes the
nonlinear transfer matrix particularly useful in dealing with
interactions whose support consists of disjoint pieces. For such an
interaction we can obtain the transfer matrix for the restriction of
the interaction to each of the disjoint pieces of its support and
then use the composition property of the transfer matrix to
determine the transfer matrix of the interaction. To demonstrate
this feature of the transfer-matrix formulation of nonlinear
scattering theory we use it to address the scattering problem for a
general double-$\delta$-function potential.

\vspace{6pt} {\em Acknowledgements} I am grateful to Neslihan Oflaz
for reading the first draft of this article and helping me
find and correct a number of minor errors and typos. This work has
been supported by the Scientific and Technological Research Council
of Turkey (T\"UB\.{I}TAK) in the framework of the project no:
114F357, and by the Turkish Academy of Sciences (T\"UBA).

\ed